\documentclass[aps,prd,nofootinbib,showpacs,preprintnumbers,amssymb,11pt]{revtex4-1} 
\pdfoutput=1
\usepackage{graphicx}
\usepackage{epsfig}
\usepackage{dcolumn}
\usepackage{bm}
\usepackage{amsmath}

\newcommand{\eqnref}[1]{Eq.~(\ref{eqn:#1})}
\newcommand{\eqnsref}[2]{Eqs.~(\ref{eqn:#1}) and (\ref{eqn:#2})}
\newcommand{\eqnsrefthree}[3]{Eqs.~(\ref{eqn:#1}),~(\ref{eqn:#2}) and (\ref{eqn:#3})}
\newcommand{\secref}[1]{Sec.~\ref{sec:#1}}
\newcommand{\secsref}[2]{Secs.~\ref{sec:#1} and \ref{sec:#2}}
\newcommand{\secsrefthree}[3]{Secs.~\ref{sec:#1},~\ref{sec:#2}  and ~\ref{sec:#3}}
\newcommand{\subsecref}[1]{Subsec.~\ref{subsec:#1}}

\newcommand{\appref}[1]{Appendix~\ref{sec:#1}}
\newcommand{\figref}[1]{Fig.~\ref{fig:#1}}

\begin{document}

\preprint{FERMILAB-PUB-12-116-T}
\preprint{NUHEP-TH/12-03}

\title{Effects from New Colored States and the Higgs Portal on Gluon
  Fusion and Higgs Decays}

\author{Kunal Kumar}
\email{kkumar@u.northwestern.edu}
\affiliation{Northwestern University, 2145 Sheridan Road, 
Evanston, IL 60208, USA}
\author{Roberto Vega-Morales}
\email{robertovegamorales2010@u.northwestern.edu}
\affiliation{Northwestern University, 2145 Sheridan Road, 
Evanston, IL 60208, USA}
\author{Felix Yu}
\email{felixyu@fnal.gov}
\affiliation{Theoretical Physics Department, Fermilab, 
Batavia, IL 60510, USA}


\begin{abstract}
We study effects from new colored states and the Higgs portal on gluon
fusion production.  We isolate possible loop contributions from new
colored scalars, fermions, and vectors, incorporating effects from
Higgs portal-induced scalar mixing, thus leading to dramatic effects
on gluon fusion and branching fractions.  Higgs identification must
generally allow for these effects, and using our results, possible
tensions from fits to the Standard Model expectation can be relieved
by inclusion of New Physics effects.
\end{abstract}


\maketitle

\section{Introduction}
\label{sec:Introduction}

The ATLAS~\cite{:2012gk} and CMS~\cite{:2012gu} experiments at the
Large Hadron Collider (LHC) have recently presented results that
indicate the observation (also supported by evidence coming from the
1.96 TeV run of the Tevatron~\cite{TEVNPH:2012ab}) of a new
resonance. When interpreted as a Standard Model (SM) Higgs, the
combined channels are consistent with the SM expectation at $1
\sigma$, yet individual channels show deviations from the SM
expectation in the 1--2$\sigma$ range.  The discovery of the Higgs
boson would certainly be one of the most exciting developments in
particle physics to date and it is tempting assume this new resonance
is indeed the Higgs, but establishing the true nature of this excess
as the SM Higgs must still proceed with due diligence.

The two main theoretical inputs in performing such a Higgs
identification are the Standard Model branching fractions for each of
the decay channels used in the combination and the overall Higgs
production cross section.  We highlight that Higgs production from
gluon fusion, the dominant production mode at hadron colliders, occurs
via loops of SM quarks and is hence uniquely sensitive to New Physics
(NP) effects arising from new colored states~\cite{Spira:1995rr,
  Binoth:1996au, DiazCruz:2000yi, Manohar:2006gz, Muhlleitner:2006wx,
  Djouadi:2007fm, Bonciani:2007ex, Arnesen:2008fb, Low:2009di,
  Bouchart:2009vq, Boughezal:2010kx, Azatov:2010pf, Boughezal:2011mh,
  Ruan:2011qg, Burgess:2009wm, Bai:2011aa, Dobrescu:2011aa,
  Belanger:2012zg, Ilisie:2012cc, Carena:2012fk, Azatov:2012rj,
  Azatov:2011qy, Berger:2012ec} or more general Higgs
portal~\cite{Schabinger:2005ei, Barbieri:2005ri, Patt:2006fw,
  Bowen:2007ia, Englert:2011yb, Cheung:2011aa, Arnold:2009ay,
  Djouadi:2011aa, Batell:2011pz, LopezHonorez:2012kv, Cohen:2012zz,
  Englert:2012ha, Djouadi:2012zc} and scalar mixing
effects~\cite{Ignatiev:2000yy, Gupta:2011gd, Frank:2012nb,
  Heckman:2012nt, deSandes:2011zs}.  The $\gamma \gamma$ branching
fraction, which arises at loop-level in the Standard
Model~\cite{Shifman:1979eb}, is similarly sensitive to NP
effects~\cite{Kribs:2007nz, Carena:2011aa, Christensen:2012ei,
  Casagrande:2010si, Goertz:2011hj}.  For example, in the well-studied
four generation Standard Model (SM4), gluon fusion rates are enhanced
while the branching ratio to diphotons is
suppressed~\cite{Arik:2002ci, Kribs:2007nz, Schmidt:2009kk, Li:2010fu,
  Anastasiou:2011qw, Passarino:2011kv, Denner:2011vt, Guo:2011ab,
  He:2011ti, Djouadi:2012ae, Kuflik:2012ai, Belotsky:2002ym}.

Moreover, the landmark discovery of a Standard Model Higgs crucially
relies on affirming the hypothesis of the Higgs mechanism for
spontaneous breaking of $SU(2)_L \times U(1)_Y$ gauge symmetry, which
predicts the existence of the Higgs boson.  This fundamental test can
only come after directly measuring the Higgs couplings to the $SU(2)$
gauge bosons.  The fact that the Higgs is responsible for chiral
symmetry breaking in the Standard Model and hence gives fermion masses
is only a byproduct of the Higgs mechanism, and thus, in particular,
the gluon fusion Higgs production mode does not directly probe the
Higgs mechanism.  This implies that NP could still be hiding in the
gluon fusion process without effecting EWSB.

This crucial point, which has also been emphasized in several recent
papers~\cite{Bai:2011aa, Cheung:2011aa, Batell:2011pz,
  Dobrescu:2011aa, Englert:2012ha}, means that the LHC Higgs searches
can be skewed by the presence of new colored particles which
positively or negatively contribute to gluon fusion.  Moreover, the
excess in the data should be interpreted not only in the context of a
SM Higgs, but also in the more exciting scenario of a possible new
scalar state which arises from a Higgs portal-induced mixing between
the SM Higgs and a new scalar.  We demonstrate that extended color
sectors involving new colored particles will generally give rise to
both effects.  In particular, if the new colored particles do not get
their mass from the vacuum expectation value (vev) of the Higgs boson,
then a generic Higgs portal term can give rise to Higgs mixing. We can
see that direct Higgs coupling and Higgs portal-induced scalar mixing
are two important categories of NP contributions that can have marked
effects on Higgs collider signals, and thus we consider them
simultaneously.

Motivated by the possibility of probing new colored states via gluon
fusion, we adopt a building block approach for an arbitrary NP model.
Namely, we isolate and calculate the gluon fusion amplitude for new
colored scalars, fermions, and vectors.  In the case of the colored
vector, we present the calculation in the context of the
renormalizable coloron model (ReCoM)~\cite{Simmons:1996fz, Hill:2002ap,
  Bai:2010dj} such that the concomitant effects from maintaining UV
consistency can be readily included.  We also allow for Higgs mixing,
where the SM Higgs is mixed with a new scalar. In addition, for a mild
and well-motivated set of assumptions, we give generic expressions for
branching ratios of the scalar mass eigenstates into the most
sensitive SM Higgs decay modes.

Since our work has some overlap with many studies in the literature,
we survey several representative papers and elaborate on the
differences.  A few recent papers have focused on our first category
of NP effects for $gg \rightarrow h$ in which the NP states couple
directly to the SM Higgs.  In particular, the authors
of~\cite{Cheung:2011aa} focused solely on the situation where new
particle masses arise from the Higgs vev, which simultaneously
sharpens their discussion of resulting gluon fusion and diphoton decay
phenomenology and limits the breadth of their conclusions.
Separately, the authors of~\cite{Dobrescu:2011aa} focused on Higgs
portal phenomenology with new colored scalars, while the work
in~\cite{Arnold:2009ay, Batell:2011pz} also included new scalars
transforming under the full SM gauge symmetry.  A similar study,
emphasizing the constraints from electroweak precision fits, was
performed in~\cite{Burgess:2009wm, Bai:2011aa}.  We go beyond these
direct coupling studies by also including the effects of a colored
vector.

There have also been a number of recent studies~\cite{Plehn:2012iz,
  Low:2012rj} where fits are performed in order to determine how
consistent the data is with a SM Higgs hypothesis.  These studies find
that generally the excess is largely consistent with a SM Higgs with a
tantalizing, but small enhancement in the $\gamma \gamma$ channel.
However, these fits still have large uncertainties, and the next data
set can change the picture drastically.  Also, the Higgs couplings to
bosons rely heavily on the vector boson fusion channel, which has
large fluctuations between 7 and 8 TeV. These uncertainties leave room
for modifications to the coupling of the SM Higgs to gluons which can
be either enhanced or suppressed given the sign of the Higgs portal
term.  We will examine this in detail below.

Regarding our second category of NP effects, when the SM Higgs mixes
with a new scalar via a Higgs portal term, a majority of the
literature has focused on the case where the New Physics sector is
completely invisible to the SM~\cite{Schabinger:2005ei,
  Barbieri:2005ri, Patt:2006fw, Bowen:2007ia, Djouadi:2011aa, 
  Djouadi:2012zc, LopezHonorez:2012kv},
providing a possible connection to the dark matter.  In this
situation, as we will see in~\secref{HiggsPortal}, only a simple
mixing angle is needed to parametrize the effects on Higgs
phenomenology, if no new decays are kinematically allowed.  Our
 work considers the more complicated scenario where
the new scalar couples to new colored particles, similar
to~\cite{Batell:2011pz}, as mentioned above.

In addition to these renormalizable scenarios of NP effects on
loop-induced SM Higgs phenomenology, a few papers have followed an
effective field theory approach by constructing and constraining the
size of dimension-six operators.  In~\cite{Manohar:2006gz}, the authors
focused on the coefficients and constraints of operators for $h
\rightarrow \gamma \gamma$, $\gamma Z$, and $gg$,
while~\cite{Low:2009di} extended the discussion to include $h
\rightarrow f \bar{f}$ as well.  Importantly, both of these
studies assume any New Physics contributions are heavy enough to be
integrated out, thus there are no new particles in the low energy
spectrum.

In contrast to the previous literature, therefore, we discuss the
general case using renormalizable interactions when both categories of
NP effects are present.  We isolate contributions with new colored
scalars, new colored fermions, including Standard Model quark mixing,
and new colored vectors, and we allow such effects to be modified by
Higgs mixing.

The paper is organized as follows.  In~\secref{HiggsPortal}, we
discuss general aspects of the Higgs portal relevant for our analysis
of Standard Model Higgs production from gluon fusion.  In~\secref{SM},
we briefly review the leading order $gg\rightarrow h$ calculation for
the Standard Model as well as the trivial extension of adding a fourth
generation.  In~\secref{Scalar}, we discuss gluon fusion in the
presence of a new colored scalar. In~\secref{Fermion}, we present the
analogous calculation for a general new colored fermion.  Lastly,
in~\secref{Vector}, we discuss the interesting case of a new colored
vector and its effects on gluon fusion in the context of a
UV-complete, renormalizable model.  Details of this calculation are
presented in~\appref{Vector_explicit}.  We summarize and conclude
in~\secref{Conclusion}.


\section{The Higgs Portal and Higgs Mixing}
\label{sec:HiggsPortal}

In this section, we review the Higgs portal as a general framework for
studying the connection between arbitrary New Physics models and Higgs
physics, with a special emphasis on the resulting effect on gluon
fusion.

In the SM, the Higgs field is responsible for breaking $SU(2)_L \times
U(1)_Y$ gauge symmetry, resulting in masses for the $W^\pm$ and $Z$
bosons as well as the chiral SM fermions.  By virtue of being the only
scalar field present in the SM, the Higgs also generates $H^\dagger
H$, which is the lowest mass dimension operator possible in the SM
that is both gauge and Lorentz invariant.  Hence, arbitrary NP
operators can then be tacked on to $H^\dagger H$ to give
\begin{equation}
\label{eqn:LOhp}
\mathcal{L}_{hp} \supset \lambda_{hp} \mathcal{O}_{NP} H^\dagger H \ .
\end{equation}
Although $\mathcal{O}_{NP}$ can be an arbitrarily high dimension
operator, with an appropriate power suppression from a high scale
$\Lambda_{NP}$, a generic Higgs portal term is only typically
unsuppressed when $\mathcal{O}_{NP}$ itself is dimension two and gauge
and Lorentz invariant: hence, we take $\mathcal{O}_{NP} \sim
\Phi^\dagger \Phi$.  One exception
is the case when a new scalar field is a pure SM and NP gauge singlet,
but since we are focused on NP effects on gluon fusion, we will not
discuss the gauge singlet case further.

One class of NP effects on gluon fusion arises from new colored states
that directly enter the $gg \rightarrow h$ loop diagram.  The direct
coupling of colored states to the Higgs via~\eqnref{LOhp} implies the
mass of the new state is shifted after electroweak symmetry breaking
(EWSB), and as this direct coupling is turned off, the NP effect
vanishes.  This class of effects is typified by models with new
colored scalars, but a new fermion with Yukawa-like couplings to the
SM Higgs boson also follows this scheme, albeit not via the Higgs
portal.  Although the case where the mass of the NP state arises
primarily from the Higgs vev was discussed in~\cite{Cheung:2011aa}, in
our more general framework the NP mass scale and the new couplings to
the SM Higgs are independent.

Since new particle masses do not have to arise from the SM Higgs vev,
a second broad class of NP effects on gluon fusion emerges.  Namely,
if a new scalar field obtains a vev to spontaneously break a new gauge
symmetry and if a Higgs portal term is present, this new scalar field
will mix with the SM Higgs.  In this class, NP effects coming from new
colored states can infiltrate gluon fusion through the mixing induced
from the Higgs portal even if these states do not directly couple to
the SM Higgs.  These effects will also exhibit the familiar
non-decoupling features in SM $gg \rightarrow h$ loop calculations by
chiral fermions or $h \rightarrow \gamma \gamma$ loop calculations by
$W$ bosons if the analogous NP states are
present~\cite{Shifman:1979eb, Shifman:2011ri, Marciano:2011gm}:
however, this non-decoupling feature only applies to the new scalar
field component of the scalar mass eigenstates.

As mentioned in the introduction, we allow for both direct and Higgs
mixing mediated categories of NP effects to be present simultaneously.
These effects arise in many extended color sector models, and we
consider isolated new colored scalars, fermions, and vectors in turn.
For colored scalars, we couple them to the Higgs via the Higgs portal
in~\eqnref{LOhp}, and hence they will exhibit an example of the direct
category of NP effects with $\lambda_{hp}$ as the direct coupling.
For colored fermions, we consider two subcategories distinguished by
the possibility of SM fermion mixing.  If new fermions are introduced
that mix with SM fermions, the usual SM calculation is modified to
accommodate fermion mass eigenstates that do not typically couple with
the SM Higgs with the usual Yukawa strength.  Without such fermion
mixing, the SM calculation is unchanged and the new contribution
arises from direct Yukawa couplings to the Higgs, the new scalar, or
both.  The decoupling behavior of new colored fermions are
parametrized by fermion mixing angles and the possible scalar mixing
angle.

Perhaps the most interesting case is that of a massive colored vector
boson.  Here, in order to have a theory which is tree level
unitary~\cite{Cornwall:1974km}, it is natural to consider an extended
color symmetry which is then spontaneously broken to $SU(3)_c$ gauge
symmetry.  Then the massive vectors corresponding to the broken
generators form representations of the unbroken color symmetry.  We
are thus left with a renormalizable, unitary, spontaneously broken
gauge theory~\cite{Fujikawa:1972fe}.

We remark that another class of New Physics effects via the Higgs
portal operator is possible.  Broadly speaking, at the renormalizable
level, where $\mathcal{O}_{NP} \sim \Phi^\dagger \Phi$
in~\eqnref{LOhp}, one class of Higgs portal effects is characterized
by new colored scalars which do not obtain vevs.  The second class is
driven by new uncolored scalars that do obtain vevs from their scalar
potential.  Another possibility is colored scalars that do obtain
vevs, but such color-breaking vacua are not viable phenomenologically.
The last possibility consists of new uncolored scalars that do not
obtain vevs from their scalar potential.  Such a scalar does not enter
the $gg \rightarrow h$ loop, but if $\lambda_{hp}$ is large and
positive, the resulting Higgs portal-induced shift in mass squared,
$-\lambda_{hp} v_h^2 / 2$ ($v_h$ is the Higgs vev) could drive the new
scalar to acquire a vev.  Hence, this last category of portal
symmetry breaking models is unique because the Higgs portal coupling
is a necessary ingredient for driving the new scalar to obtain a
nonzero vev.  Obviously, the roles of the new scalar and the Higgs
scalar can be reversed, whereby the Higgs portal term allows a new
scalar vev to drive the Higgs field to obtain a negative mass squared
and hence trigger EWSB.  We reserve a study of ``Portal Symmetry
Breaking'' phenomenology for future work.  Also, in the discussion
above, we have delineated cases according to specific constraints on
the Lagrangian parameters.  A precise determination of these bounds
would require an analysis of renormalization group evolution, which is
beyond the scope of this work.

\subsection{New Physics Scalar -- Standard Model Higgs Mixing}
\label{subsec:mixing}
We briefly discuss the second class of NP effects from the Higgs
portal described above, {\it i.e.} a new scalar and the SM Higgs both
obtain vevs in~\eqnref{LOhp} and mix.  For simplicity, we only
consider one new scalar, but our discussion is readily generalized to
multiple scalars.  We also assume $\Phi$ transforms as a singlet under $SU(2)_L\times U(1)_Y$, but that it is charged under a new local or global symmetry in order to prevent "tadpole" terms.  We let $\mathcal{O}_{NP} \sim \Phi^\dagger \Phi$
for a new scalar field $\Phi$, giving
\begin{equation}
\label{eqn:Lhp}
\mathcal{L} \supset 
\lambda_{hp} H^\dagger H \Phi^\dagger \Phi 
\sim \lambda_{hp} v_h v_\phi h \phi \ ,
\end{equation}
where we have suppressed representation indices and expanded the
fields $H \sim \dfrac{1}{\sqrt{2}} (h + v_h)$ and $\Phi \sim
\dfrac{1}{\sqrt{2}} (\phi + v_{\phi})$.  We assume the scalar
potentials $V(\Phi)$ and $V(H)$ are also present and~\eqnref{Lhp} is
the only Lagrangian term involving both $\Phi$ and $H$ fields.  The
usual stability, triviality, and renormalizability constraints on the
full scalar potential $V(H) + V(\Phi) - \lambda_{hp} |H|^2 |\Phi|^2$
are assumed to be satisfied and will be imposed when we consider
explicit models in~\secsref{Fermion}{Vector}.  Here, since $\Phi$
obtains a vev,~\eqnref{Lhp} leads to mixing via the mass matrix
\begin{equation}
\label{eqn:scalar_mass_matrix}
m_{\text{scalar}}^2 = \left( \begin{array}{cc}
m_h^2 & -\lambda_{hp} v_h v_{\phi} \\
-\lambda_{hp} v_h v_{\phi} & m_{\phi}^2 \\
\end{array} \right) \ ,
\end{equation}
where $v_h$ and $v_{\phi}$ are calculated from minimizing the full
potential $V(H) + V(\Phi) - \lambda_{hp} |\Phi|^2 |H|^2$ and hence
determine $m_h$ and $m_{\phi}$.  The functional dependence of $m_h$
and $m_{\phi}$ on their respective potential parameters can be fixed
by solving the potentials $V(H)$ and $V(\Phi)$ separately, and in the
limit that $\lambda_{hp} \rightarrow 0$, the exact vevs $v_h$ and
$v_{\phi}$ recover their original, unperturbed values.  This
observation has important ramifications when calculating the exact
Goldstone--Goldstone--scalar couplings needed for vector loop
amplitudes in Feynman gauge, which are discussed
in~\subsecref{GoldstoneSector}.

We can readily diagonalize the symmetric mass
matrix~\eqnref{scalar_mass_matrix} to obtain the mass eigenstates
\begin{equation}
\label{eqn:s1s2_definition}
\begin{array}{ccc}
s_1 &=& h \cos \theta - \phi \sin \theta \ , \\
s_2 &=& h \sin \theta + \phi \cos \theta \ , \\
\end{array}
\end{equation}
with a Jacobi rotation mixing angle $\theta$ defined by
\begin{equation}
\label{eqn:mixingangle}
\tan 2\theta = \dfrac{ -2 \lambda_{hp} v_h v_{\phi}}{m_{\phi}^2 -
  m_h^2} \ .
\end{equation}
We will also need the inverse operations,
\begin{equation}
\label{eqn:hphi_definition}
\begin{array}{ccc}
h    &=&  s_1 \cos \theta + s_2 \sin \theta \ , \\
\phi &=& -s_1 \sin \theta + s_2 \cos \theta \ . \\
\end{array}
\end{equation}
The eigenvalues of~\eqnref{scalar_mass_matrix} are
\begin{equation}
\label{eqn:Ms1}
m^2_{s_1} = \dfrac{1}{2} \left( m_h^2 + m_\phi^2 \right) -  
\dfrac{1}{2} \sqrt{ \left( -m_h^2 + m_\phi^2 \right)^2 + 4 \lambda_{hp}^2 v_h^2
  v_\phi^2 } \ ,
\end{equation}
and
\begin{equation}
\label{eqn:Ms2}
m^2_{s_2} = \dfrac{1}{2} \left( m_h^2 + m_\phi^2 \right) +
 \dfrac{1}{2} \sqrt{ \left( -m_h^2 + m_\phi^2 \right)^2 + 4 \lambda_{hp}^2 v_h^2
  v_\phi^2 } \ ,
\end{equation}
where we have taken $m_{s_1} < m_{s_2}$ without loss of generality.
As mentioned before and demonstrated in~\cite{Schabinger:2005ei,
  Bowen:2007ia, Batell:2011pz}, the mixing of the scalar states from
the Higgs portal can significantly affect scalar production via gluon
fusion.  Moreover, the mixing is driven purely by the strength of
$\lambda_{hp}$, which must be real but whose sign is not fixed.

\subsection{New Physics Effects on Production of $s_{1,2}$}
\label{subsec:general_amplitudes}

We can now readily disentangle the two categories of New Physics
effects on gluon fusion.  Now, because of $h$--$\phi$ mixing via the
Higgs portal in~\eqnref{Lhp}, we must calculate cross sections for $gg
\rightarrow s_1$ and $gg \rightarrow s_2$ production instead of the
gauge eigenstates $h$ and $\phi$.  Since both $h$ and $\phi$ can
couple to new colored particles, contributions to $gg \rightarrow
s_{1,2}$ can manifest themselves through both the $h$ and $\phi$
components of $s_{1,2}$, leading to suppression or enhancement of the
production rate relative to the SM.  This also implies that partial
decay widths are affected, whereas in hidden sector models, such
widths are unaltered apart from a universal $\cos^2 \theta$
suppression coming from Higgs mixing.

From the discussion above, we can decompose the production amplitude
of $s_1$ via gluon fusion in terms of the gauge eigenstate $h$ and
$\phi$ production amplitudes as,
\begin{equation}
\label{eqn:general_s1_s2_amps}
\begin{array}{ccc}
\mathcal{M} (gg \rightarrow s_1) &=& 
\left. 
c_\theta \left[ \mathcal{M} (gg \rightarrow h) \right]
\right|_{m_h = m_{s_1}}
\left. 
-s_\theta \left[ \mathcal{M} (gg \rightarrow \phi) \right]
\right|_{m_\phi = m_{s_1}} \\
\mathcal{M} (gg \rightarrow s_2) &=& 
\left.
 s_\theta \left[ \mathcal{M} (gg \rightarrow h) \right]
\right|_{m_h = m_{s_2}}
\left.
+c_\theta \left[ \mathcal{M} (gg \rightarrow \phi) \right]
\right|_{m_\phi = m_{s_2}} \ , \\
\end{array}
\end{equation}
where $c_\theta \equiv \cos \theta$, $s_\theta \equiv \sin \theta$ are
defined by~\eqnref{mixingangle}.  In the discussion below, we presume
the matrix elements are evaluated at the appropriate scalar mass and
will drop the notation above.  Hence, given the linear combination
dictated by~\eqnref{general_s1_s2_amps}, we are now free to isolate
the contributions to $gg \rightarrow h$ and $gg \rightarrow \phi$.

We are particularly interested in identifying, at the amplitude level,
the mechanisms responsible for modifying gluon fusion and whether and
how they can decouple.  A completely general expression for all
possible NP effects along these lines is cumbersome, so instead we
write
\begin{equation}
\label{eqn:general_s1_sfv_amp}
\begin{array}{ccl}
\vspace{4pt} 
\mathcal{M} (gg \rightarrow s_1) &=& 
c_\theta \left[ \mathcal{M} (gg \xrightarrow[scalars]{} h) + 
\mathcal{M} (gg \xrightarrow[fermions]{} h) +
\mathcal{M} (gg \xrightarrow[vectors]{} h) \right] \\
&-& s_\theta \left[ \mathcal{M} (gg \xrightarrow[scalars]{} \phi) + 
\mathcal{M} (gg \xrightarrow[fermions]{} \phi) +
\mathcal{M} (gg \xrightarrow[vectors]{} \phi) \right] \ , \\
\end{array}
\end{equation}
and treat each category of loop particles separately.\footnote{For the
  vector loop calculation, we implicitly assume a unitary gauge
  calculation where only vectors appear in the loop.  If working in
  Feynman gauge, the associated Goldstone and ghosts would also be
  part of the vector category.}  Each of these categories can be
further subdivided into particles that couple solely to $h$, solely to
$\phi$, or simultaneously to both.  In the scalar case, for example,
we can write
\begin{equation}
\label{eqn:general_s1_scalar_amp}
\begin{array}{ccl}
\vspace{4pt} 
\mathcal{M} (gg \xrightarrow[scalars]{} s_1) &=& 
c_\theta \left[ 
\sum\limits_i \mathcal{M} (gg \xrightarrow[\eta_i]{} h) +
\sum\limits_j \mathcal{M} (gg \xrightarrow[\eta_j]{} h) \right] \\
&-& s_\theta \left[ 
\sum\limits_j \mathcal{M} (gg \xrightarrow[\eta_j]{} \phi) +
\sum\limits_k \mathcal{M} (gg \xrightarrow[\eta_k]{} \phi) \right]
\ , \\
\end{array}
\end{equation}
where the scalars $\eta_i$, $\eta_j$, $\eta_k$ couple only to $h$,
both to $h$ and $\phi$, and only to $\phi$, respectively.  We can now
make definitive statements about the decoupling behavior of the
scalars $\eta_i$, $\eta_j$ and $\eta_k$.  If the masses of $\eta_i$
($\eta_k$) arise solely from the vev $v_h$ ($v_\phi$), then these
scalars will exhibit non-decoupling from $h$ ($\phi$) as their masses
are taken very large: if instead their masses include sources besides
$v_h$ or $v_{\phi}$, then decoupling will occur as the mass scale of these 
new sources is taken large.  The behavior of the
$\eta_j$ states are a straightforward combination of the previous
arguments.

For fermions, we write
\begin{equation}
\label{eqn:general_s1_fermion_amp}
\begin{array}{ccl}
\vspace{4pt} 
\mathcal{M} (gg \xrightarrow[fermions]{} s_1) &=& 
c_\theta \left[ 
\sum\limits_i \mathcal{M} (gg \xrightarrow[\psi_i]{} h) +
\sum\limits_j \mathcal{M} (gg \xrightarrow[\psi_j]{} h) \right] \\
&-& s_\theta \left[ 
\sum\limits_j \mathcal{M} (gg \xrightarrow[\psi_j]{} \phi) +
\sum\limits_k \mathcal{M} (gg \xrightarrow[\psi_k]{} \phi) \right]
\ . \\
\end{array}
\end{equation}
To be more illustrative, we can take some familiar examples to
demonstrate the flexibility of~\eqnref{general_s1_fermion_amp}.  In
the case with Higgs mixing but without new fermions $\psi_j$ or
$\psi_k$, then $\psi_i$ consists of the SM quarks and we get a
universal $c_\theta$ suppression of the matrix element.  If instead we
only add a new vector-like top partner to the SM, then $c_\theta = 1$,
$s_\theta = 0$, and $\psi_i$ includes the first five SM quarks and the
two fermion mass eigenstates resulting from top mixing while the
$\psi_j$ and $\psi_k$ sums are absent.  Finally, if Higgs mixing is
present and new colored fermions are added that couple both to $h$ and
$\phi$ but do not mix with the SM fermions, then $\psi_i$ will run
over the SM quarks and $\psi_j$ will run over the NP colored fermions.

Lastly, we can introduce massive colored vectors.  We will only
consider the case where these vectors couple to $\phi$, giving the
relatively simple expression
\begin{equation}
\label{eqn:general_s1_vector_amp}
\begin{array}{ccl}
\mathcal{M} (gg \xrightarrow[vectors]{} s_1) &=&
-s_\theta \left[ 
\sum\limits_k \mathcal{M} (gg \xrightarrow[V_k]{} \phi) \right]
 \ ,
\end{array}
\end{equation}
emphasizing that this contribution to the gluon fusion rate for $s_1$
production relies on the Higgs portal, since the SM Higgs is assumed
to play no role in breaking the extended color gauge symmetry.

After the above discussion, we present a parametric understanding of
how production and decays of $s_{1,2}$ are affected by direct coupling
and $h$--$\phi$ mixing.  As we have seen, performing a completely
general analysis would be overly cumbersome, and so we will make a few
mild assumptions to make the analysis more intuitive and tractable.
Throughout the discussion, we assume a narrow width approximation,
allowing us to factorize production and decay processes.

We define the overall leading order enhancement or suppression factor
of $s_1$ production relative to SM Higgs production via gluon fusion
as
\begin{equation}
\label{eqn:epsilon_definition}
\epsilon_{gg} \equiv \dfrac{\sigma (gg \rightarrow s_1)}{\sigma (gg
  \xrightarrow[SM]{} h)} = \dfrac{ \left| \mathcal{M} (gg \rightarrow s_1)
  \right|^2 }{ \left| \mathcal{M} (gg \xrightarrow[SM]{} h) \right|^2 }
= \dfrac{ \left|
       c_\theta \mathcal{M} (gg \rightarrow h) 
     - s_\theta \mathcal{M} (gg \rightarrow \phi) \right|^2}{
\left| \mathcal{M} (gg \xrightarrow[SM]{} h) \right|^2 }
= c^2_\theta\left| \mathcal{Z}_{ggh} 
       - t_\theta \mathcal{Z}_{gg \phi} \right|^2 \ ,
\end{equation}
using~\eqnref{general_s1_s2_amps} and with $t_\theta = \tan \theta$.
The complex amplitude ratios are given by
\begin{equation}
\label{eqn:Zgg_definition}
\mathcal{Z}_{ggh} \equiv \dfrac{\mathcal{M} (gg \rightarrow
  h)}{\mathcal{M} (gg \xrightarrow[SM]{} h)} \qquad
\mathcal{Z}_{gg\phi} \equiv \dfrac{\mathcal{M} (gg \rightarrow
  \phi)}{\mathcal{M} (gg \xrightarrow[SM]{} h)} \ ,
\end{equation}
and will simplify significantly for any given NP model under
consideration, as we will demonstrate
in~\secsrefthree{Scalar}{Fermion}{Vector}.  We see
that both $\epsilon_{gg} > 1$ (signaling enhancement) and
$\epsilon_{gg} < 1$ (signaling suppression) are possible with New
Physics and changing the sign of $\lambda_{hp}$.  In the limit that
$\theta = 0$, the only effect on gluon fusion arises from the
inclusion of new colored states that directly couple to the SM Higgs,
which was a main focus of~\cite{Cheung:2011aa, Bai:2011aa,
  Dobrescu:2011aa}.  In the case where Higgs mixing is the only new
effect, then $\mathcal{Z}_{ggh} = 1$ and $\mathcal{Z}_{gg\phi} = 0$,
and we have the simple expression $\epsilon_{gg} = c^2_\theta$, as
noted in~\cite{Schabinger:2005ei}.

We remark that complete suppression of gluon fusion does not
correspond to vanishing LHC production for the $s_1$ state.  This is
because the subdominant modes of vector boson fusion, vector boson
association, and $t \overline{t} h$ production comprise 12.5\% of the
total cross section for a SM Higgs mass at 125
GeV~\cite{Dittmaier:2011ti}.  Moreover, even if the leading order
cancellation in~\eqnref{epsilon_definition} is exact, we expect NLO
corrections, which can be as large as 20\% in the case of colored
stops~\cite{Muhlleitner:2006wx}, to make the cancellation imperfect.

\subsection{New Physics Effects on Decays of $s_{1,2}$}
\label{subsec:s12_decays}
We now extend our discussion to include NP effects on decay widths for
our scalar state $s_1$, which we take to be dominantly SM Higgs-like.
We will not detail all of the (practically infinite!) possible final
states for $s_1$, but will instead focus on the $WW$, $ZZ$, $\gamma
\gamma$, $b \overline{b}$ and $\tau^+ \tau^-$ decay channels.  For the
$WW$ final state, we write
\begin{equation}
\label{eqn:s1_MEtoWW}
\mathcal{M} (s_1 \rightarrow WW) =
 c_\theta \mathcal{M} (h \rightarrow WW)
-s_\theta \mathcal{M} (\phi \rightarrow WW)
\approx c_\theta \mathcal{M} (h \rightarrow WW) \ ,
\end{equation}
and thus
\begin{equation}
\label{eqn:s1_BRtoWW}
\dfrac{\mathcal{B}(s_1 \rightarrow WW)}{\mathcal{B}
(h \xrightarrow[SM]{} WW)} 
\approx c_\theta^2 \dfrac{\Gamma_h}{\Gamma_{s_1}} \ ,
\end{equation}
where we have assumed the tree-level coupling of $h WW$ dominates the
(typically loop-induced) coupling of $\phi WW$, and $\Gamma_h$ and
$\Gamma_{s_1}$ are the total width of the purely SM Higgs and the mass
eigenstate $s_1$, respectively.  Under the same assumption that $h ZZ$
dominates the $\phi ZZ$ coupling, the same result
in~\eqnref{s1_BRtoWW} also applies to the $ZZ$ final state, and so
branching ratios of $s_1$ to $WW$ or $ZZ$ diboson states are typically
suppressed in Higgs mixing models.

For the diphoton final state, we can adapt our gluon fusion
discussion, replacing colored particles with electromagnetically
charged particles. Following the guide of~\eqnref{general_s1_s2_amps},
this gives
\begin{equation}
\label{eqn:s1_MEtogammagamma}
\begin{array}{ccc}
\mathcal{M} (s_1 \rightarrow \gamma \gamma) =
\left. c_\theta \left[ \mathcal{M} (h \rightarrow \gamma \gamma) \right]
\right|_{m_h = m_{s_1}}
- \left. s_\theta \left[ 
\mathcal{M}(\phi \rightarrow \gamma \gamma) \right] 
\right|_{m_\phi = m_{s_1}} \ .
\end{array}
\end{equation}
Unlike the $WW$ or $ZZ$ decay modes, the $h \rightarrow \gamma \gamma$
decay is induced at loop level in the SM and new contributions can
easily cancel against or add to the SM contributions.
Using~\eqnref{s1_MEtogammagamma}, we can write the relative branching
ratio as,
\begin{equation}
\label{eqn:s1_BRtogammagamma}
\dfrac{ \mathcal{B}(s_1 \rightarrow \gamma \gamma)}{
       \mathcal{B}(h \xrightarrow[SM]{} \gamma \gamma)} = 
\epsilon_{\gamma\gamma} \dfrac{\Gamma_h}{\Gamma_{s_1}} \ ,
\end{equation}
where $\epsilon_{\gamma\gamma}$ is analogous to $\epsilon_{gg}$
in~\eqnref{epsilon_definition} and $\mathcal{Z}_{h\gamma\gamma}$ and
$\mathcal{Z}_{\phi\gamma\gamma}$ are defined similarly.

The relative rate for $gg \rightarrow s_1 \rightarrow \gamma \gamma$
versus $gg \rightarrow h \rightarrow \gamma \gamma$ is now given by
\begin{equation}
\label{eqn:totalrate}
\mathcal{R} = \epsilon_{gg}\epsilon_{\gamma\gamma} 
\dfrac{\Gamma_h}{\Gamma_{s_1}} \ .
\end{equation}
In many models, though, the various inputs for~\eqnref{totalrate}
reduce to simple expressions.  For example, in Higgs mixing scenarios
where $\phi$ only couples to hidden sector particles, we obtain
$\mathcal{Z}_{h\gamma \gamma} = 1$, $\mathcal{Z}_{\phi \gamma \gamma}
= 0$, and so
\begin{equation}
\label{eqn:s1_BRtogammagamma2}
\dfrac{ \mathcal{B}(s_1 \rightarrow \gamma \gamma)}{
       \mathcal{B}(h \xrightarrow[SM]{} \gamma \gamma)} = 
c_\theta^2 
\dfrac{\Gamma_h}{\Gamma_{s_1}} \ ,
\end{equation}
which agrees with the universal $c_\theta^2$ suppression noted
in~\cite{Schabinger:2005ei}.  Another simple limiting case arises if
we take $\theta = 0$ and introduce new charged particles in the
$\gamma \gamma$ loop coupling to the Higgs.  In this case, $h \equiv
s_1$ and we can write
\begin{equation}
\label{eqn:s1_BRtogammagamma3}
\dfrac{ \mathcal{B}(s_1 \rightarrow \gamma \gamma)}{
       \mathcal{B}(h \xrightarrow[SM]{} \gamma \gamma)} =
\dfrac{\Gamma_h}{\Gamma_{s_1}}
\left| \mathcal{Z}_{h\gamma\gamma} \right|^2  \ ,
\end{equation}
so that only the direct NP effects contribute.

Finally, we can calculate the $s_1$ branching ratio to $b \bar{b}$ or
$\tau^+ \tau^-$.  If Higgs mixing is present, if $\phi$ does not
appreciably couple to the SM fermions, and if the SM fermions are not
mixed with NP fermions, then the same results from~\eqnref{s1_BRtoWW}
apply, substituting $f \bar{f}$ for $WW$.  A completely general
expression, however, because of the possible presence of all of these
effects, is unwieldy.  As an explicit case, for the $b \bar{b}$ final
state, if we allow for $h$--$\phi$ mixing and introduce a 
coupling between $\phi$ and $b \bar{b}$, we obtain
\begin{equation}
\label{eqn:s1_MEtoffbar}
\dfrac{ \mathcal{B}(s_1 \rightarrow b\bar{b})}{
       \mathcal{B}(h \xrightarrow[SM]{} b\bar{b})} = 
c_\theta^2 \dfrac{\Gamma_h}{\Gamma_{s_1}}
\left| 1 - t_\theta \mathcal{Z}_{\phi b\bar{b}} \right|^2 \ ,
\end{equation}
where $\mathcal{Z}_{\phi b\bar{b}} = \mathcal{M} (\phi \rightarrow b
\bar{b}) / \mathcal{M} (h \xrightarrow[SM]{} b \bar{b})$.  We can see that
interference effects from $\mathcal{Z}_{\phi b\bar{b}}$, although
diluted by $t_\theta$, can lead to an overall increase in the
branching fraction of $s_1 \rightarrow b \bar{b}$.

In summary, we have disentangled the effects from Higgs portal-induced
mixing of $h$ and $\phi$ from NP effects caused by direct coupling to
$h$, $\phi$, or both.  For gluon fusion, we have explicitly identified
the decoupling behavior of New Physics states
in~\eqnsrefthree{general_s1_scalar_amp}{general_s1_fermion_amp}{general_s1_vector_amp}.
If we assume NP couplings to be small or negligible, then the
resulting $s_1$ branching ratio has a universal $c_\theta^2$
suppression and a universal total width ratio suppression.  On the
other hand, interference effects resulting from couplings to $h$
and/or $\phi$ lead to a myriad of effects and possibilities for both
suppression and enhancement of relative rates can be achieved.

We note that all of these expressions can readily be adapted for $s_2$
decay with an appropriate $c_\theta \rightarrow s_\theta$, $-s_\theta
\rightarrow c_\theta$ exchange and $m_{s_1} \rightarrow m_{s_2}$.  In
addition, if $m_{s_2} > 2 m_{s_1}$, there is the additional decay mode
$s_2 \rightarrow s_1 s_1$, as emphasized in~\cite{Bowen:2007ia}.
Also, if any of the new states are lighter than $m_{s_1} /2$ or
$m_{s_2} / 2$, then additional non-standard decay modes open up. This
effect is manifest in the above expressions through the ratio of total
widths $\Gamma_h / \Gamma_{s_1}$.

\section{The $gg\rightarrow h$ Process in SM}
\label{sec:SM}
\begin{figure}[htb]
\begin{center}
\includegraphics[scale=0.75, angle=0]{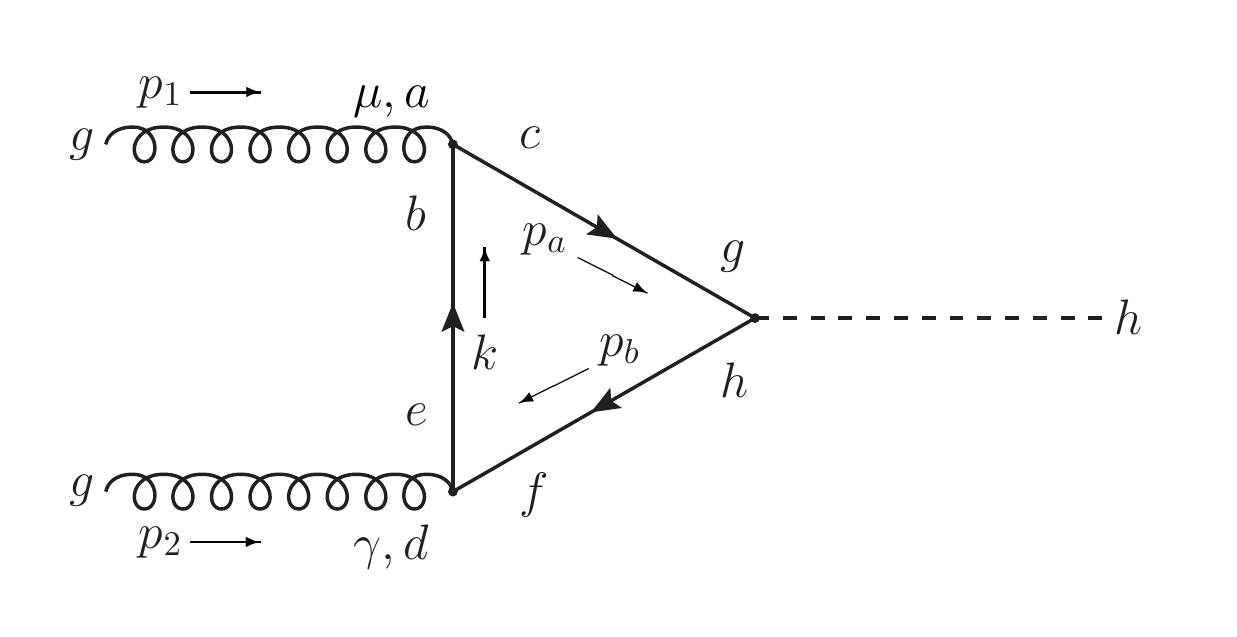}  
\caption{\label{fig:SM_ggh}{The Standard Model contribution
    to $gg \rightarrow h$. }}
\end{center}
\end{figure}

Here we briefly review the leading order Standard Model calculation
for Higgs production via gluon fusion.  As shown in~\figref{SM_ggh},
gluon fusion arises in the SM via quark loops, with the dominant
contribution coming from the top quark with its large Yukawa coupling.
We again highlight the fact that since neither the $W$ or $Z$ boson
couplings are probed in this production mode, large effects can be
present in this loop process that strongly change Higgs production but
do not affect EWSB.

The total Standard Model amplitude is
\begin{equation}
\label{eqn:fermionamp}
i \mathcal{M}_{SM}^{ad} = \sum\limits_f i \mathcal{M}_f^{ad} 
= \sum\limits_f i \left( \dfrac{\alpha_s}{\pi} \right) 
\dfrac{C(r_f)}{2v_h} \delta^{ad} \epsilon_{1\mu} \epsilon_{2\nu} 
\left( p_1^\nu p_2^\mu - \dfrac{m_h^2}{2} g^{\mu \nu} \right) F_F(\tau_f) \ ,
\end{equation}
where $f$ runs over the SM quarks, $C(r_f)$ is the Casimir invariant
($C(r_f) = 1/2$ for SM quarks), $a$ and $d$ are color indices, $p_1
\cdot p_2 = \dfrac{m_h^2}{2}$ for an on-shell Higgs, $\tau_f \equiv
m_h^2 / (4 m_f^2)$ and $F_F(\tau)$ is the well-known loop function
\begin{equation}
\label{eqn:F_F}
F_F(\tau) = \dfrac{2}{\tau^2} \left( \tau + (\tau - 1) 
Z(\tau) \right) \ ,
\end{equation}
using
\begin{equation}
\label{eqn:Z_SM}
Z(\tau) = \left\{ \begin{array}{c}
\arcsin^2 \sqrt{\tau} \qquad  \tau \leq 1 \\
\dfrac{-1}{4} \log \left[ 
\dfrac{ 1 + \sqrt{1 - \tau^{-1}} }{
        1 - \sqrt{1 - \tau^{-1}} } - i \pi 
\right]^2 \qquad \tau > 1 \ . \\
\end{array} \right.
\end{equation}
Because the SM quarks obtain their mass purely from EWSB, they do not
decouple even as we take the limit $\tau \rightarrow 0$ (equivalent to
$m_f \rightarrow \infty$).  In the case of the SM4, this sum would
include $t'$ and $b'$ contributions.  In the limit that the SM Higgs
is well below the threshold for $t$, $t'$, and $b'$ to propagate
on-shell in~\figref{SM_ggh}, we obtain the usual factor of 3
enhancement in the SM4 matrix element for $gg \rightarrow h$, which
gives, at leading order, a factor of 9 enhancement for gluon fusion
production in SM4 relative to SM3~\cite{Denner:2011vt}.  Recent
literature that has attempted to resolve the quandary of a sequential
fourth generation of fermions with the lack of enhancement in gluon
fusion include Refs.~\cite{Schmidt:2009kk, Guo:2011ab, He:2011ti,
  Djouadi:2012ae, Kuflik:2012ai}.

We can anticipate, in the presence of new additions to gluon fusion
coming from New Physics, that the main structure
of~\eqnref{fermionamp} will not change apart from possible differences
in the scalar vertex, $C(r)$, and the loop function $F(\tau)$.  In
particular, the $p_1^\nu p_2^\mu - p_1 \cdot p_2 g^{\mu \nu}$
structure of the matrix element is assured by $SU(3)_c$ gauge
invariance (or equivalently, the Ward identity).  This is analogous to
the situation in the $h \rightarrow \gamma \gamma$ calculation, where
electromagnetic gauge invariance requires the same momentum
structure~\cite{Marciano:2011gm}.

\section{New Colored Scalar}
\label{sec:Scalar}

In this section, we isolate and calculate the effect of a colored
complex scalar $S$ propagating in the $gg \rightarrow h$ loop.  We use
the Higgs portal in~\eqnref{Lhp} to couple $S$ to the SM Higgs, and we
write a (positive) tree-level mass squared for $S$ such that $SU(3)_c$
remains unbroken and Higgs mixing is absent.  Depending on the sign
and strength of $\lambda_{hp}$, we can achieve significant suppression
or enhancement of gluon fusion as a result of the interference between
the SM fermions and the colored scalar.

The Lagrangian involving $S$ is
\begin{equation}
\label{eqn:ScalarL}
\mathcal{L}_S = \left|D_\mu S \right|^2 - m_0^2 S^\dagger S
- \kappa |S^\dagger S|^2 
+ \lambda_{hp} S^\dagger S H^\dagger H \ ,
\end{equation}
where color indices have been suppressed and we assume $m_0^2 > 0$ and
$\kappa > 0$ to ensure stability.  As discussed
in~\secref{HiggsPortal}, $\lambda_{hp}$ must be real: for positive
(negative) $\lambda_{hp}$, we will get destructive (constructive)
interference with the SM loop calculation, in agreement
with~\cite{Dobrescu:2011aa, Batell:2011pz} (note we have a different
sign convention for $\lambda_{hp}$).  After EWSB, the physical scalar
mass is
\begin{equation}
\label{eqn:mS_definition}
m_S^2 \equiv m_0^2 - \frac{1}{2} \lambda_{hp} v_h^2 \ ,
\end{equation}
which imposes the constraint that $m_0^2 > \frac{1}{2} \lambda_{hp} v_h^2$ to
avoid portal symmetry breaking of $SU(3)_c$.

\begin{figure}[htb]
\includegraphics[width=0.45\textwidth]{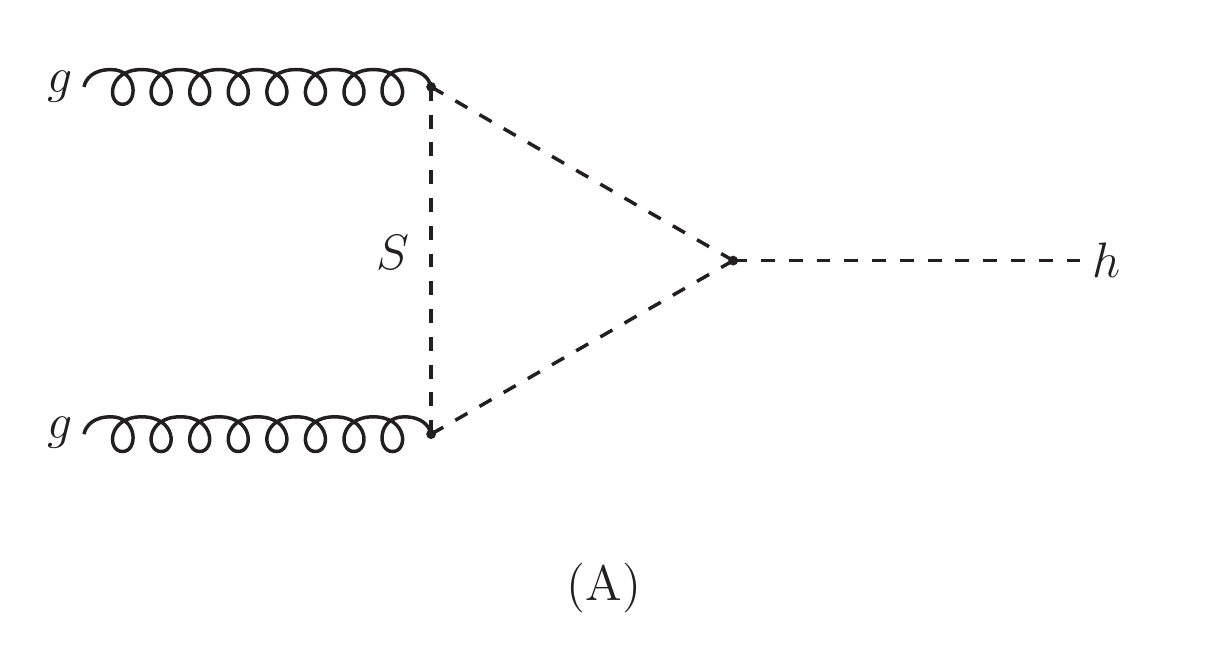}  
\includegraphics[width=0.45\textwidth]{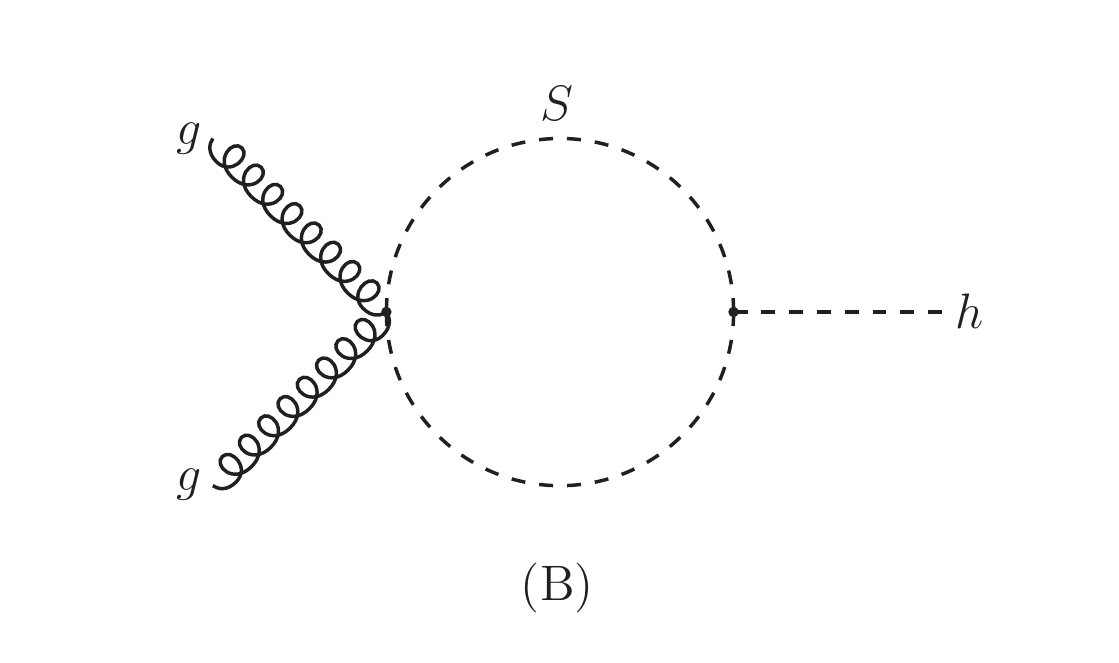}  
\caption{\label{fig:scalarloop}{Feynman diagrams for scalar loop
    contributions to $gg \rightarrow h$.  For a complex scalar one
    must also include the charge conjugate equivalent of diagram
   (A).}}
\end{figure}

The two diagrams to calculate are shown in~\figref{scalarloop}.  Since
$S$ is complex, the matrix element for~\figref{scalarloop}A needs to
be multiplied by 2 to account for the charge conjugate diagram: if $S$
were real, no factor of 2 is used and instead the matrix element
for~\figref{scalarloop}B must include a symmetry factor of (1/2).

The total amplitude corresponding to the diagrams
in~\figref{scalarloop} for a complex scalar field propagating in the
loop is
\begin{equation}
\label{eqn:scalar_ME}
i \mathcal{M}_S^{ad} = i \left( \dfrac{\alpha_s}{\pi} \right)
\left( \dfrac{ C(r_S) \lambda_{hp} v_h }{4 m_S^2} \right)
\delta^{ad} \epsilon_{1\mu} \epsilon_{2\nu} 
(p_1^\nu p_2^\mu - \dfrac{m_h^2}{2} g^{\mu \nu}) F_S(\tau_S) \ ,
\end{equation} 
where $C(r_S)$ is the $SU(3)_c$ Casimir invariant for $S$, $\tau_S =
m_h^2 / (4 m_S^2)$ and the loop function $F_S$ is defined to be
\begin{equation}
\label{eqn:F_S}
F_S(\tau) = \tau^{-1} - \tau^{-2} Z(\tau) \ ,
\end{equation} 
with $Z(\tau)$ defined in~\eqnref{Z_SM}.  Note that for fixed
$\lambda_{hp}$ the amplitude decouples as $m_S \rightarrow \infty$
because of the tree-level mass squared $m_0^2$.

Now, the summed amplitude for $\mathcal{M} (gg \xrightarrow[SM+S]{}
h)$ is
\begin{equation}
\label{eqn:SM3+S_ME}
\begin{array}{ccl}
i \mathcal{M}_{SM+S}^{ad} &=& 
 i \left( \sum\limits_{f} \mathcal{M}_f^{ad} \right) + 
 i \mathcal{M}_S^{ad} \\
&=& i \left( \dfrac{\alpha_s}{\pi} \right) 
\delta^{ad} \epsilon_{1\mu} \epsilon_{2\nu}
(p_1^\nu p_2^\mu - \dfrac{m_h^2}{2} g^{\mu \nu}) \left( 
\sum\limits_{f} \left( \dfrac{C(r_f)}{2v_h} F_F(\tau_f) \right)
+ \dfrac{C(r_S) \lambda_{hp} v_h}{4 m_S^2} F_S(\tau_S)
\right) \ . \\
\end{array}
\end{equation}
If $m_S$, $m_t > m_h/2$, then $F_S$ is strictly real and negative 
and $F_F$ is strictly real and
positive, which implies that for $\lambda_{hp} > 0$ ($\lambda_{hp} <
0$) the interference between the colored scalar amplitude and the SM
is destructive (constructive).

Since the phase space integration needed to calculate the $s_1$ cross
section is identical to the SM Higgs case, we can write the ratio
$\epsilon_{gg}$ from~\eqnref{epsilon_definition} as
\begin{equation}
\label{eqn:SM+S_epsilon}
\left. \epsilon_{gg} \right|_{SM+S} = \dfrac{ \left| 
\sum\limits_{f} \left( \dfrac{C(r_f)}{2v_h} F_F(\tau_f) \right)
+ \dfrac{C(r_s) \lambda_{hp} v_h}{4 m_S^2} F_S(\tau_S) 
\right|^2 }{ \left|
\sum\limits_{f} \left( \dfrac{C(r_f)}{2v_h} F_F(\tau_f) \right)
\right|^2 } \ .
\end{equation}

\begin{figure}[htb]
\includegraphics[width=0.48\textwidth]{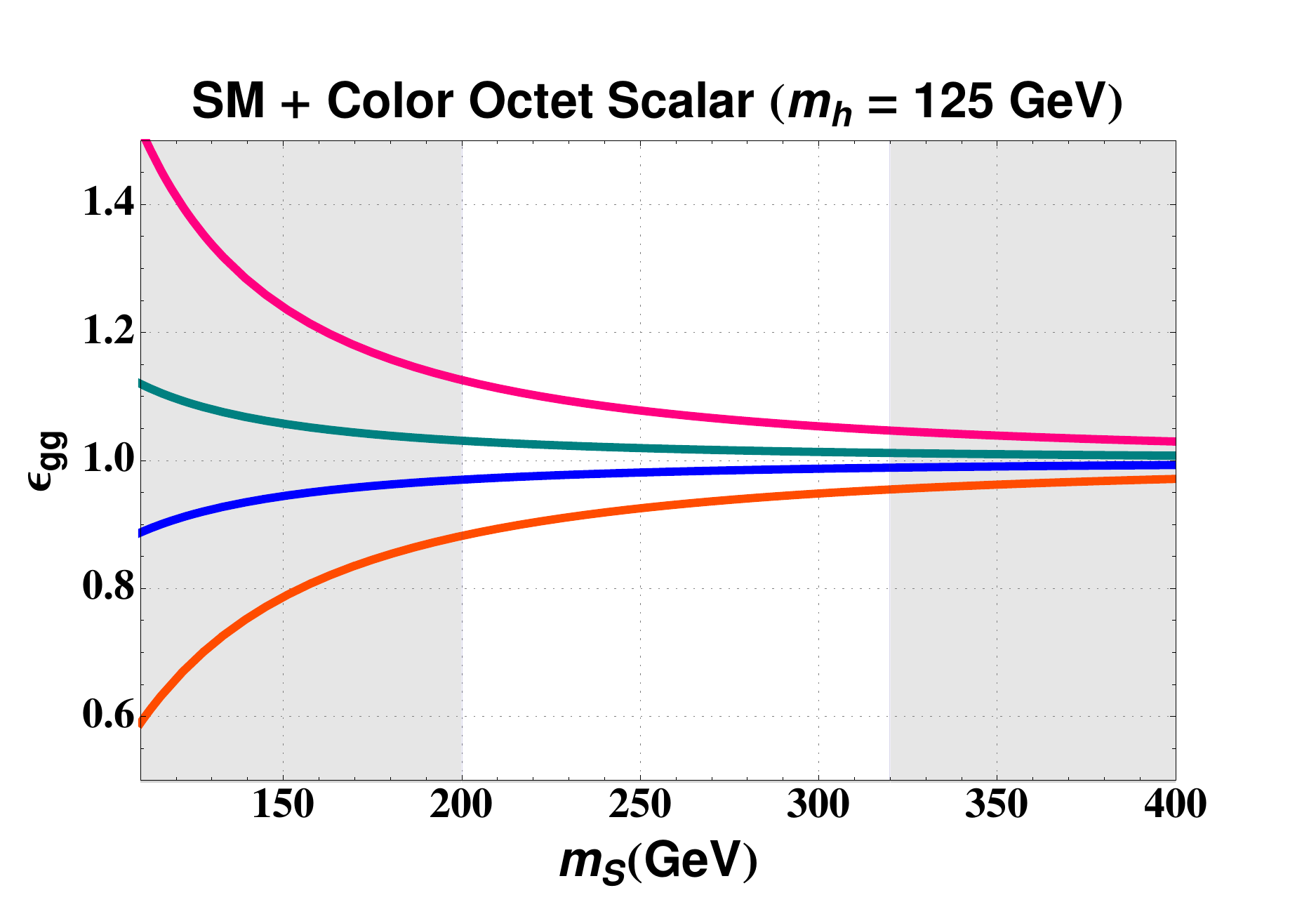}
\includegraphics[width=0.48\textwidth]{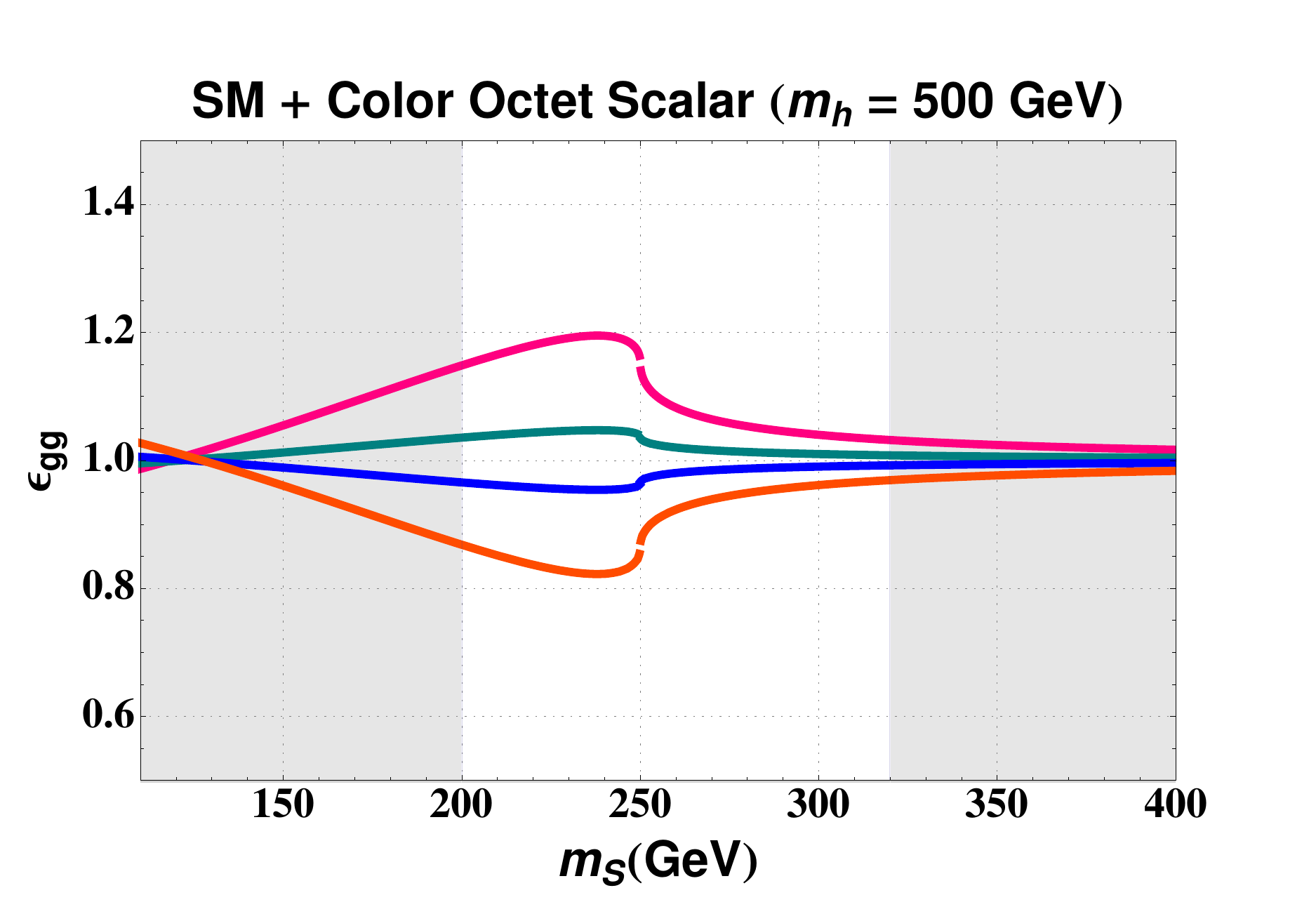}
\includegraphics[width=0.48\textwidth]{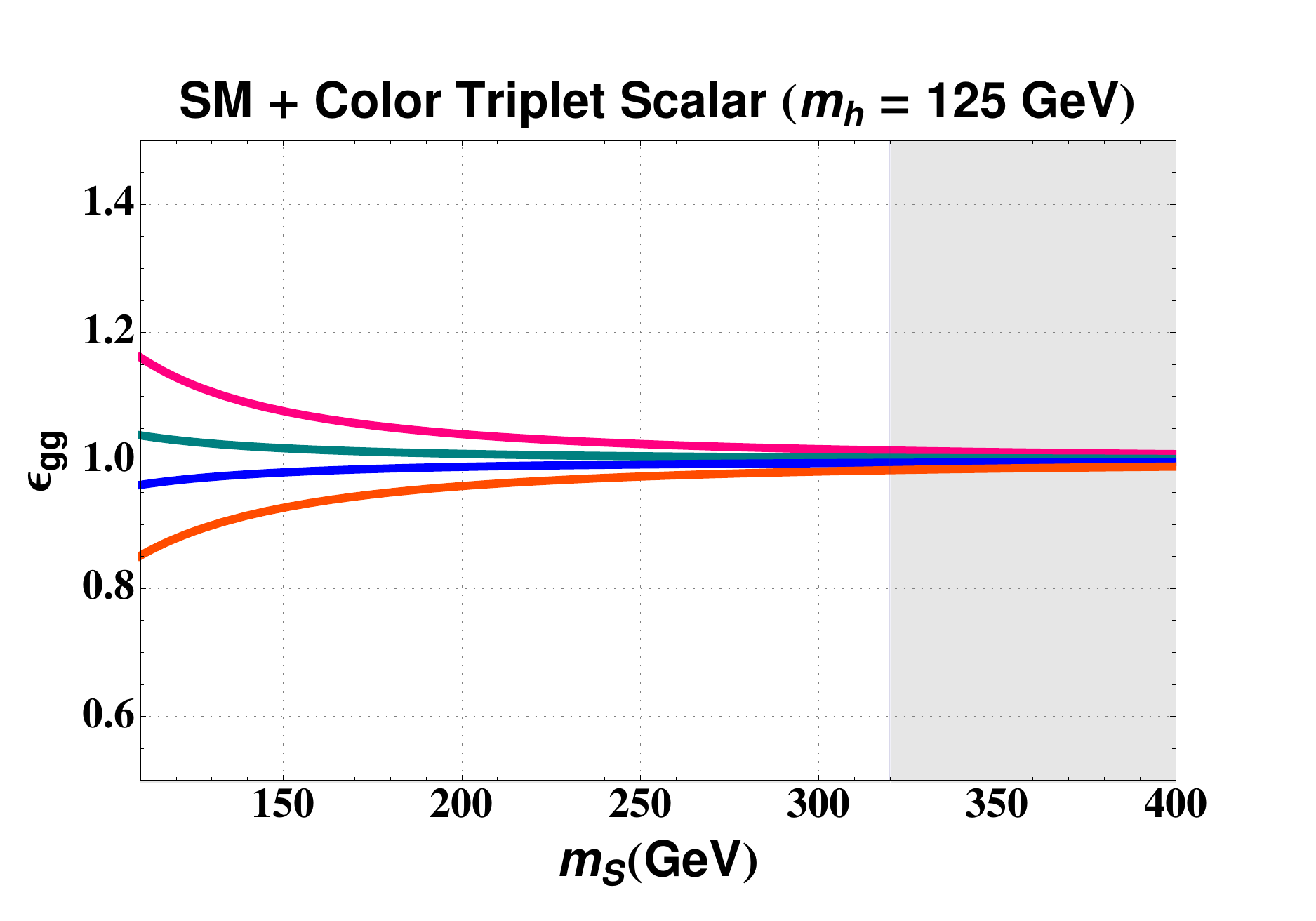}
\includegraphics[width=0.48\textwidth]{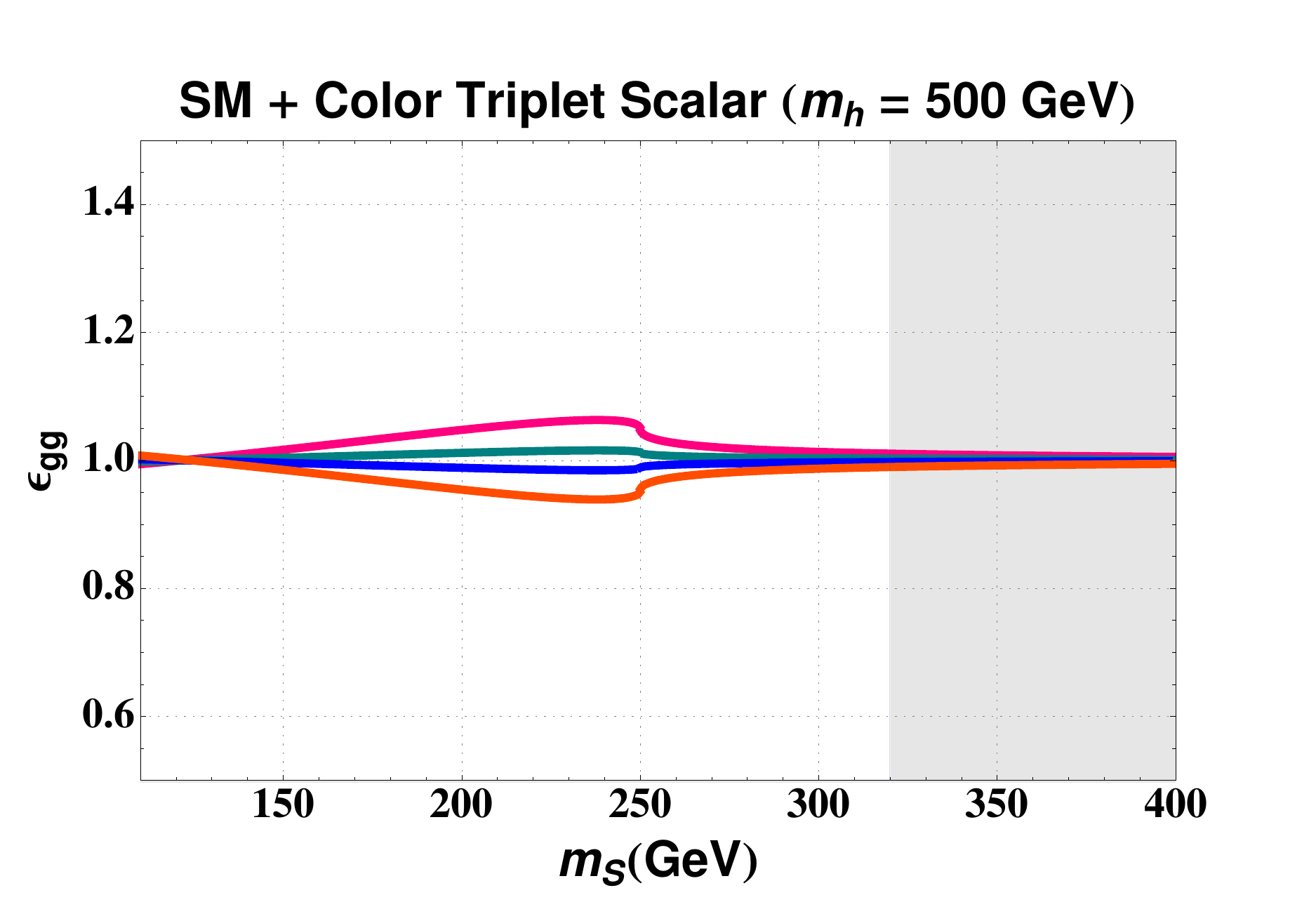}
\caption{\label{fig:SM3LMsContour_fixlhp}{Relative rate
    $\epsilon_{gg}$ in~\eqnref{SM+S_epsilon}, showing the effect of
    the inclusion of a real color octet scalar (top row) or complex
    color triplet scalar (bottom row), for $m_h = 125$ GeV (left
    panels) or $m_h = 500$ GeV (right panels).  At the center of each
    plot, from top to bottom, the solid lines correspond to
    $\lambda_{hp} =-0.1$, $-0.025$, $0.025$, $0.1$.  The left (right)
    gray bands in the octet scalar plots come from the ATLAS (CMS)
    search for pair produced dijet resonances.  For the triplet case,
    the CMS bound still applies but the ATLAS bound is unconstraining
    after rescaling cross sections. }}
\end{figure}

We consider the addition of a real color octet scalar ($C(r_s)=3$,
symmetry factor of $1/2$) and a complex color triplet scalar ($C(r_s)
= 1/2$) and plot $\epsilon_{gg}$ in~\figref{SM3LMsContour_fixlhp} as a
function of $m_S$ for some representative choices of parameters $m_h$
and $\lambda_{hp}$.  For the SM calculation, we sum over bottom and
top quark contributions, using $m_b = 4.20$~GeV and $m_t = 172.5$~GeV.
We adopt the results of~\cite{Altmannshofer:2012ur} to draw vertical
exclusion bands on light color octet scalars from
ATLAS~\cite{Aad:2011yh} and CMS~\cite{CMS-PAS-EXO-11-016} in dijet
pair resonance searches.  The gap in sensitivity from 200 GeV to 320
GeV between the 34 pb$^{-1}$ ATLAS search and the 2.2 fb$^{-1}$ CMS
search is a result of the increased multijet trigger threshold to
handle more difficult run conditions.  In particular, for the CMS
study, turn-on effects of the QCD multijet trigger made the background
modeling unreliable below 320 GeV.

We see that both enhancement and suppression are possible, typically
delineated by the choice of the sign of $\lambda_{hp}$. This is
manifest in the region where $m_S > m_h/2$ where $\lambda_{hp} > 0$
corresponds to a suppression and $\lambda_{hp} < 0$ corresponds to an
enhancement. In the region where $m_S < m_h/2$, we see enhancement and
suppression for both signs of $\lambda_{hp}$ since the scalars can go
on-shell in the loop, leading to an additional imaginary contribution
to the scalar amplitude.  The resulting interference is complicated by
our inclusion of the bottom quark and its imaginary contribution, so
the overall magnitude has competing cancellations among real and
imaginary amplitude pieces.  We note
that~\figref{SM3LMsContour_fixlhp} shows the expected decoupling of
$S$ as $m_S$ grows.  We also remark that for negative values of
$\lambda_{hp}$, our results are consistent
with~\cite{Dobrescu:2011aa}, where the finite difference in our
results is a result of our inclusion of the bottom quark.  Lastly,
with regards to the ATLAS dijet pair search, we note that the complex
triplet scalar is 1/9 the production cross section of the real octet
scalar, if their masses are equal, rendering the search insensitive to
complex triplet scalars.

\section{New Colored Fermion}
\label{sec:Fermion}
Adding new colored fermions to the Standard Model can greatly affect
gluon fusion SM Higgs production in a number of unique ways.  On one
hand, new sequential generations of chiral fermions will add
constructively with the SM fermion loops and, at leading order, scale
the top quark loop by a multiplicative factor, as discussed
in~\secref{SM}.  On the other hand, a new vector-like colored fermion
that does not mix with SM fermions has no effect on gluon fusion.  In
general, a new colored fermion mass eigenstate comprised of chiral and
vector-like components will enhance the SM Higgs gluon fusion rate
according to the chiral projection of the mass eigenstate.

Because we also allow for Higgs portal-induced scalar mixing, though,
the general situation can lead to either an overall suppression or
enhancement of the gluon fusion rate.  A model demonstrating the
myriad of competing effects is straightforward to construct but only
illuminating in its limiting cases.  Hence, we will initially consider
only mixing between a NP fermion and a SM fermion, neglecting the
Higgs portal and Higgs mixing.

We add new vector-like top partners~\cite{Choudhury:2001hs,
  Kumar:2010vx}, given by
\begin{equation}
\label{eqn:psiRrep}
\chi_{L,R} \sim \left( 3, 1 \right)_{2/3} \ .
\end{equation}
This leads to the following mass terms,
\begin{equation}
\label{eqn:masslag}
\mathcal{L} \supset -y_t \tilde{H} \bar{Q}_L t_R 
- y_L \tilde{H} \bar{Q}_L \chi_R - M \bar{\chi}_L \chi_R 
+ \text{ h.c.} \ ,
\end{equation}
where $M$ is a free parameter and $y_L$ induces mixing between the SM
top quark and $\chi$.  In the $(t, \chi)$ gauge basis, we have mass
$\hat{M}$ and interaction $\hat{N}_h$ matrices given by
\begin{equation}
\label{eqn:MN_matrices}
\hat{M} = 
\left( \begin{array}{cc}
M_t & \xi_L \\  0 & M \\
\end{array} \right) \ , \qquad
\hat{N}_h = 
\left( \begin{array}{cc}
M_t & \xi_L \\  0 & 0 \\
\end{array} \right) \ ,
\end{equation}
with $\xi_L = \dfrac{y_L v_h}{\sqrt{2}}$ and $M_t = \dfrac{y_t
  v_h}{\sqrt{2}}$.  Note the $0$ entry in $\hat{M}$ can always be
ensured since it corresponds to the $M^\prime \bar{\chi}_L t_R$
operator, which can be trivially rotated away since $\chi_R$ and $t_R$
have the same quantum numbers.  The mass basis rotation matrices are
defined in the usual way from $\hat{R} (\hat{M}^\dagger\hat{M})
\hat{R}^\dagger = \left| \hat{M}_D \right|^2$ and $\hat{L} (\hat{M}
\hat{M}^\dagger) \hat{L}^\dagger = \left| \hat{M}_D \right|^2$.  After
rotating $\hat{M}$ and $\hat{N}_h$ on the left (right) by a
left-handed (right-handed) rotation matrix, we obtain
\begin{equation}
\label{eqn:mass_basis_lagrangian}
\begin{array}{ccc}
\mathcal{L} \supset 
- \bar{\textbf{t}} 
\left( \hat{M}_D + \dfrac{h}{v_h} \hat{V}_h \right)
P_R~\textbf{t} + \text{ h.c.} \ ,
\end{array}
\end{equation}
where $\textbf{t} \equiv (t_1, t_2)$ and $\hat{M}_D = \hat{L} \hat{M}
\hat{R}^\dagger = \text{ diag}(m_{t_1}, m_{t_2})$, $\hat{V}_h =
\hat{L} \hat{N}_h \hat{R}^\dagger$.  The coupling matrix $\hat{V}_h$
dictates the couplings of the top sector to the SM Higgs and, in
principle, can have off diagonal entries; however, only the diagonal
entries contribute to $gg \rightarrow h$, because the $\hat{L}$ and
$\hat{R}$ rotations leave the gauge interactions diagonal in the mass
basis.

\begin{figure}[htb]
\begin{center}
\includegraphics[scale=0.65, angle=0]{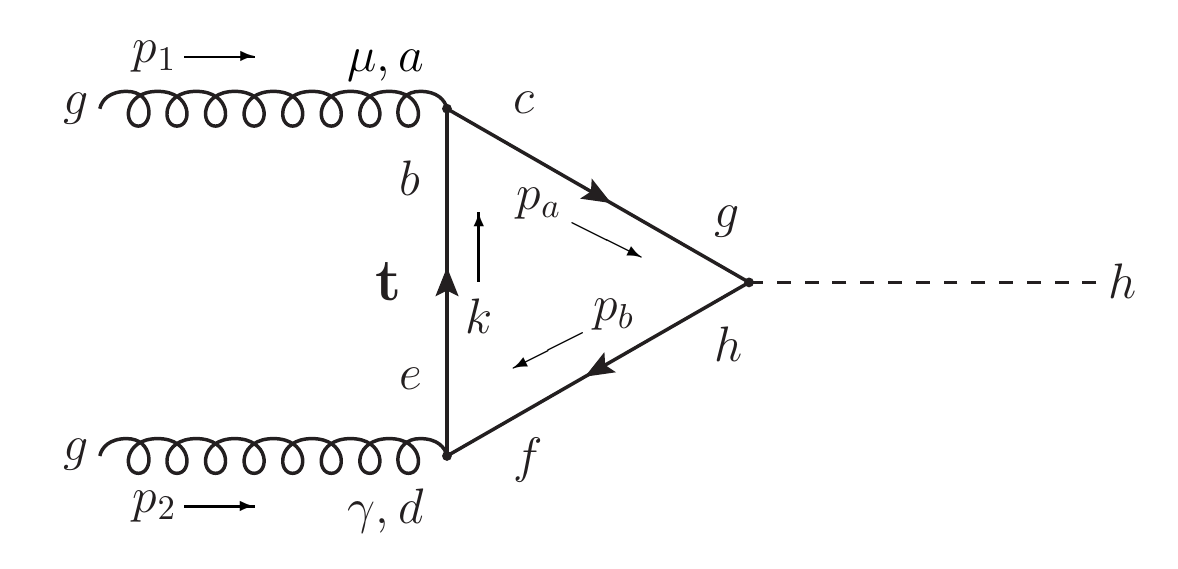}  
\caption{\label{fig:mixed_fermion_diagrams}{Exotic fermion
    contribution in the mass eigenbasis.}}
\end{center}
\end{figure}

In this top partner scenario, each mass eigenstate gives its own
contribution to the diagrams in~\figref{mixed_fermion_diagrams}.
Since $SU(3)_c$ gauge invariance guarantees these two contributions
differ only in their coupling to the Higgs via the element of
$\hat{V}_h$, we can take the SM result for $gg \rightarrow h$ through
a top quark and insert the appropriate element of $\hat{V}_h$ in place
of the usual Yukawa coupling.  This approach also encompasses more
complicated mixing scenarios whereby the matrix element will exhibit
different combinations of mixing angles and couplings for the various
$\hat{V}_h$ entries as a replacement for the $h f \bar{f}$ vertex in
the $gg \rightarrow h$ amplitude.  Since we are focused on exotic
fermion effects on $gg \rightarrow h$, we take the $(\hat{V}_h)_{ij}$
entry to be a free parameter, which can be readily calculated in any
complete model.

The amplitudes involving exotic fermions in the loop are
\begin{equation}
\label{eqn:exotic_fermionamp}
i \mathcal{M}_{F}^{ad} = i \sum\limits_i 
\left( \dfrac{\alpha_s}{\pi} \right)
\left( \dfrac{(\hat{V}_h)_{ii}}{m_{F_i}} \right) 
\left( \dfrac{C(r_{F_i})}{2v_h} \right)
\delta^{ad} \epsilon_{1\mu} \epsilon_{2\nu} 
\left( p_1^\nu p_2^\mu - \dfrac{m_h^2}{2} g^{\mu \nu} \right)
F_F(\tau_{F_i}) \ ,
\end{equation}
where the repeated indices on $(\hat{V}_h)_{ii}$ are not summed,
$F_F(\tau)$ is given by~\eqnref{F_F}, $\tau_{F_i} \equiv m_h^2 /
(4m_{F_i}^2)$, and $F_i \in \{ t_1, t_2 \}$.  We see that the
amplitude decouples as $m_{F_i} \rightarrow \infty$, unless $m_{F_i}$
and $(\hat{V}_h)_{ii}$ are generated by a common scale such as the
Higgs vev.  These direct new physics contributions will alter $gg
\rightarrow h$ even in the absence of Higgs mixing.  Generally these
contributions will add constructively if $(\hat{V}_h)_{ii} > 0$.

Now, we augment the previous discussion to include Higgs mixing
between $h$ and a new scalar $\phi$.  We replace the vector-like mass
$M$ in~\eqnref{masslag} by a Yukawa term which generates the desired
mass term and a new interaction term involving $\phi$ after $\phi$
obtains a vev, giving
\begin{equation}
\label{eqn:vectormass_generation}
y_\phi \phi \bar{\chi}_L \chi_R \Rightarrow 
M (1 + \dfrac{\phi}{v_\phi}) \bar{\chi}_L \chi_R \ ,
\end{equation}
where $M = \dfrac{y_\phi v_\phi}{\sqrt{2}}$.  We now have a second
interaction matrix $\hat{V}_\phi$, which is added
to~\eqnref{mass_basis_lagrangian} and defined analogously to
$\hat{V}_h$, where
\begin{equation}
\label{eqn:Mmatrix}
\hat{N}_\phi = 
\left( \begin{array}{cc}
0 & 0 \\ 0 & M \\
\end{array} \right) \ .
\end{equation}
As discussed in~\subsecref{general_amplitudes}, the $t_1$, $t_2$ loops
will give an enhancement factor for gluon fusion production given by
\begin{equation}
\label{eqn:SM+F_epsilon}
\begin{array}{l}
\left. \epsilon_{gg} \right|_{SM + \chi_{L,R}} = \\
\dfrac{ c^2_\theta \left| 
\sum\limits_{f, \text{ no }t} \left( \dfrac{C(r_f)}{v_h} F_F(\tau_f) \right)
+ \sum\limits_i \left( \dfrac{C(r_{F_i})}{v_h} 
\left( \dfrac{ (\hat{V}_h)_{ii}}{m_{F_i}} \right)  F_F(\tau_{F_i}) 
- t_\theta 
 \dfrac{C(r_{F_i})}{v_\phi} 
\left( \dfrac{ (\hat{V}_\phi)_{ii}}{m_{F_i}} \right)  F_F(\tau_{F_i}) \right)
\right|^2 }{ 
\left|
\sum\limits_{f} \left( \dfrac{C(r_f)}{v_h} F_F(\tau_f) \right)
\right|^2 } \ . \\
\end{array}
\end{equation}
Note that as $m_{t_1,t_2}\rightarrow \infty$, $t_1$ will decouple from the
$\phi$ component of $s_1$ but not the $h$ component, and vice versa
for $t_2$.  

\begin{figure}[htb]
\begin{center}
\includegraphics[width=0.48\textwidth, angle=0]{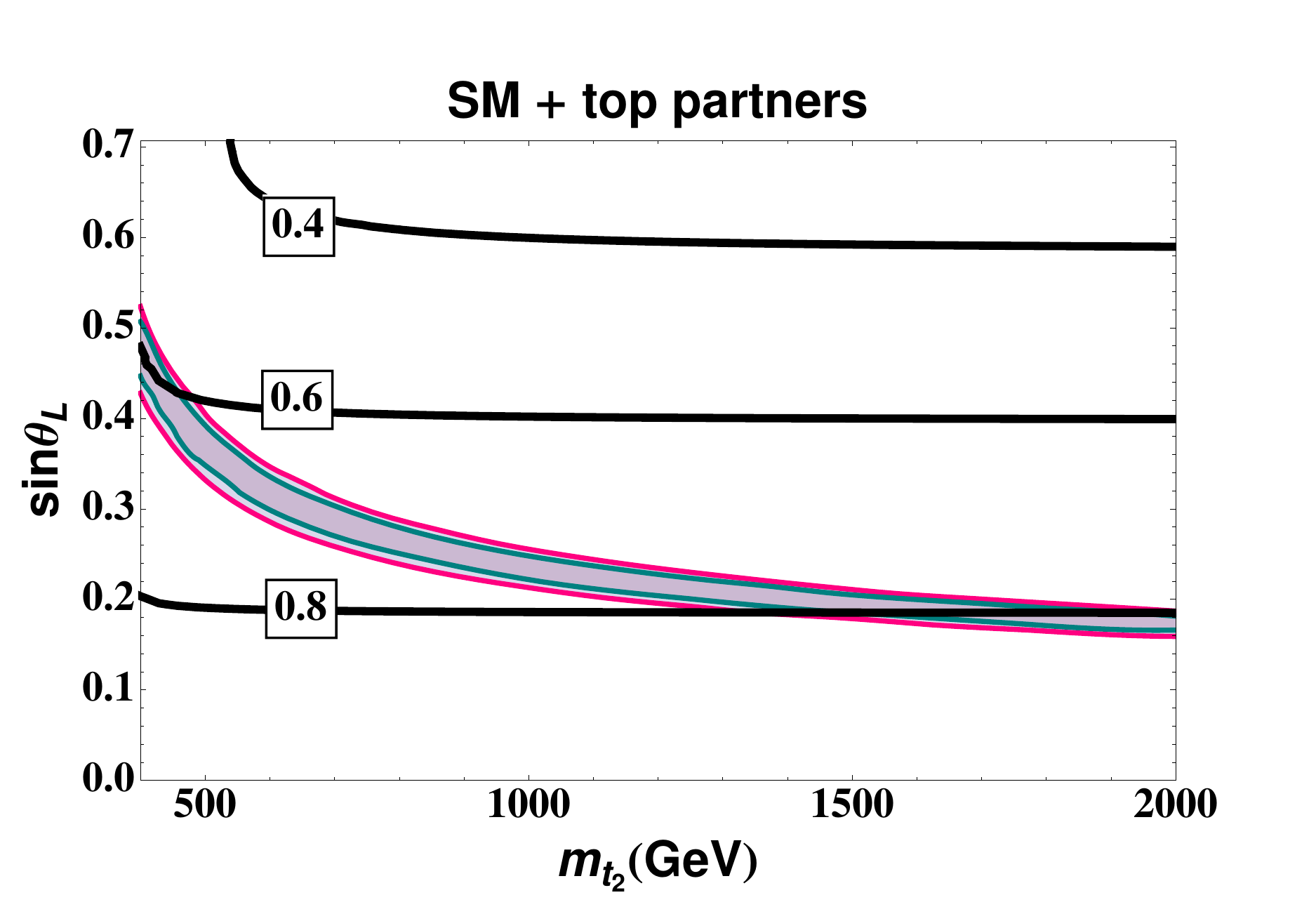}  
\includegraphics[width=0.48\textwidth, angle=0]{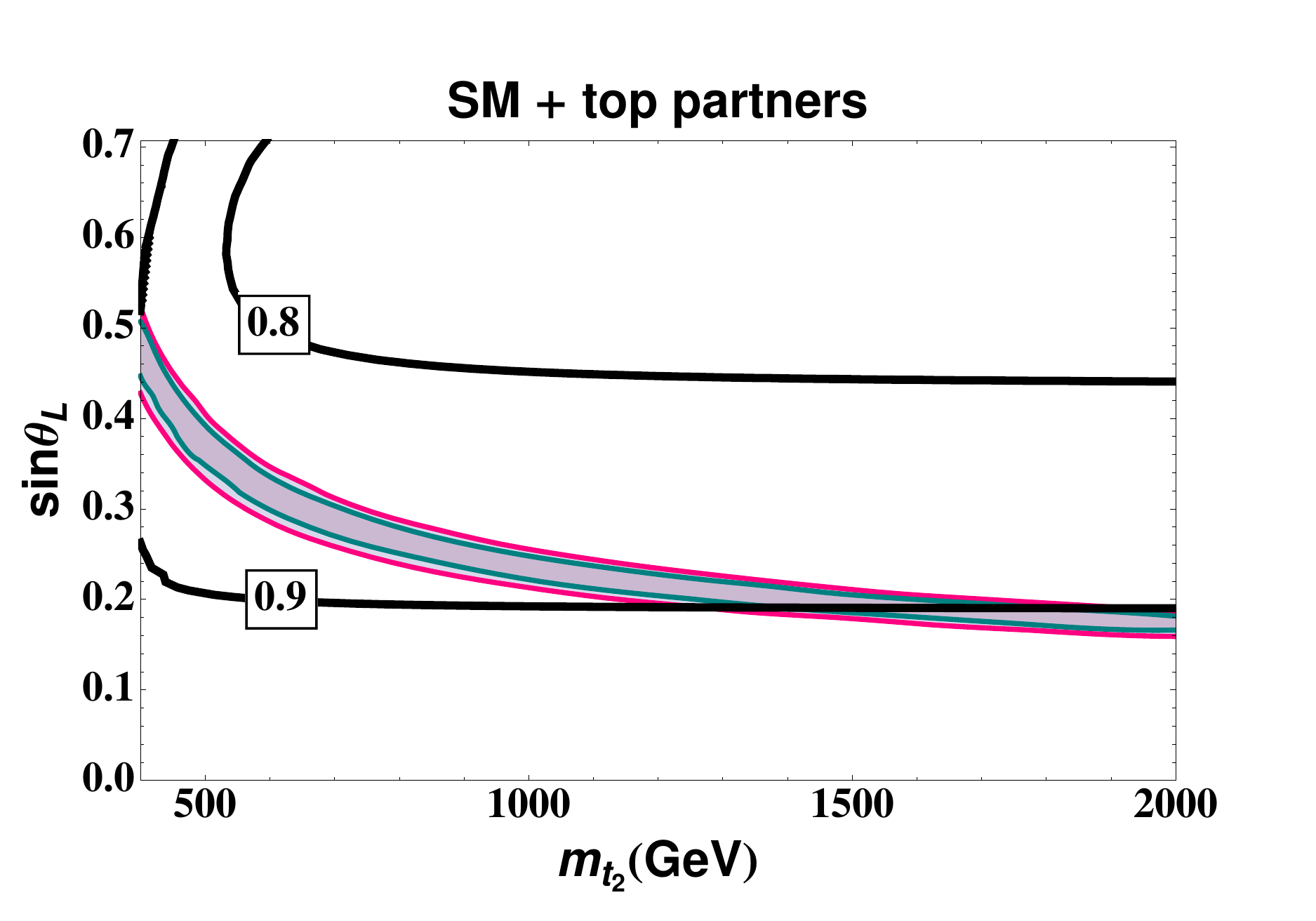}  
\caption{\label{fig:mixed_fermion_relative_rate}{Contours of the
    relative rate of $s_1$ production as a function of $m_{t_2}$ and
    the left-handed mixing angle in the top partner scenario for
    $m_{s_1} = 800$ GeV, $v_\phi = 500$ GeV, and $\lambda_{hp} = -1$
    (left) and $0.75$ (right).  We adapt the analysis and results of
    Ref.~\cite{Bai:2011aa} to show shaded contours which are
    consistent with the oblique parameters $S$ and $T$ at the 68\% and
    95\% C.L.. }}
\end{center}
\end{figure}

We recognize that these new top partners will induce shifts in the
electroweak oblique parameters $S$ and $T$, but a full analysis of the
allowed top partner parameter space is beyond the scope of this study.
Therefore, we adapt the results from Ref.~\cite{Bai:2011aa}, which
studied top partner effects on Higgs production and included the
constraints from the $S$ and $T$ oblique corrections.  We set
$m_\phi^2 > m_h^2$, and for the vevs we are considering,
$-\lambda_{hp} v_h v_\phi$ is a small perturbation to the diagonal
mass terms in~\eqnref{scalar_mass_matrix}, allowing us to approximate
the $s_1$ contribution to $S$ and $T$ by the Higgs contribution
considered in~\cite{Bai:2011aa} for equal masses.  We can thus
illustrate our main point, suppression of gluon fusion, in this
phenomenologically viable top partner scenario.
In~\figref{mixed_fermion_relative_rate} we plot contours of
$\epsilon_{gg}$ as a function of the left-handed fermion mixing angle
and the heavy fermion mass eigenstate, $m_{t_2}$ for representative
values of $\lambda_{hp}$.  The shaded bands correspond to regions
consistent with the oblique parameters at the 68 \% and 95 \% C.L.,
taken from~\cite{Bai:2011aa}.

\begin{figure}[htb]
\begin{center}
\includegraphics[width=0.48\textwidth, angle=0]{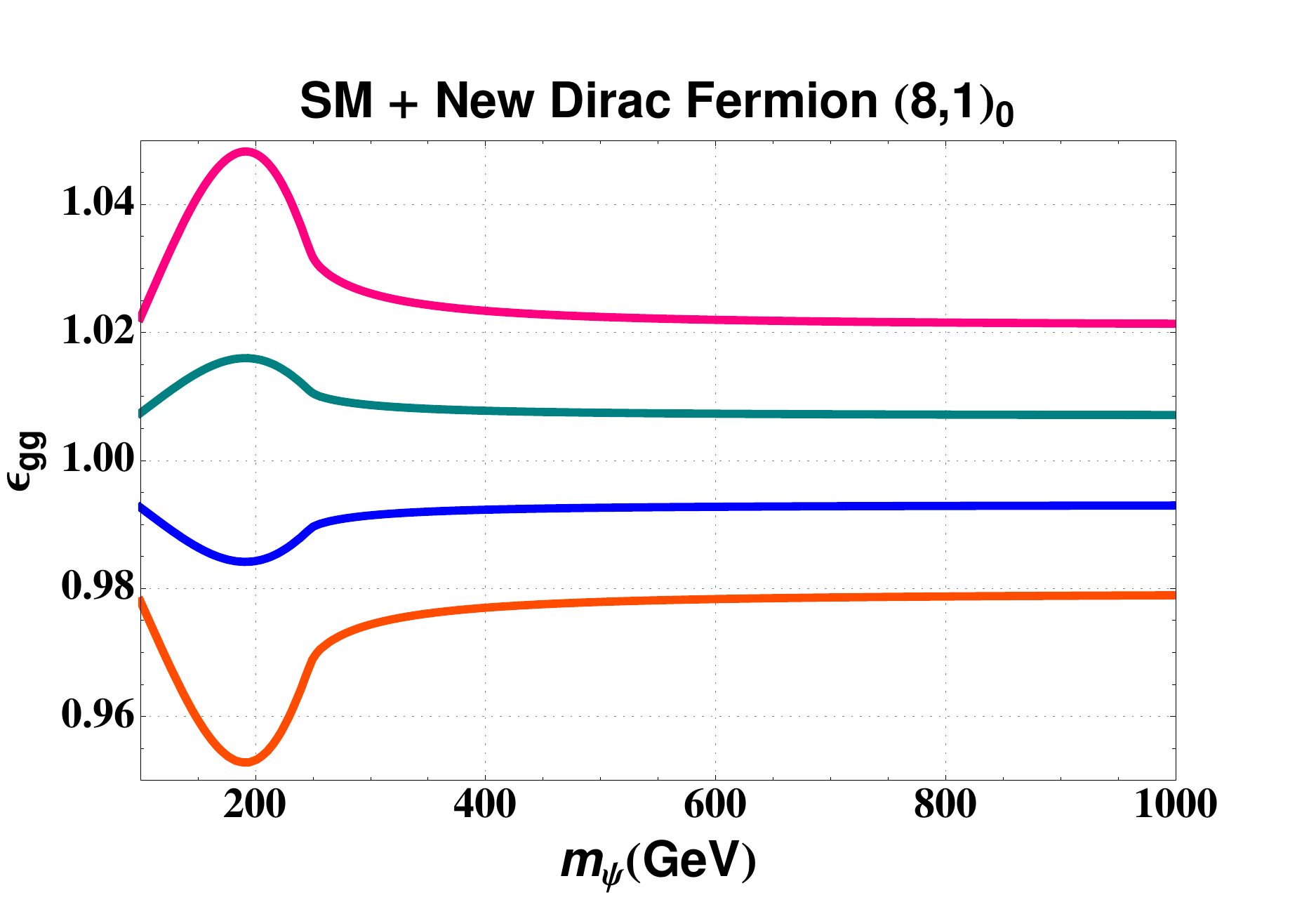}  
\includegraphics[width=0.48\textwidth, angle=0]{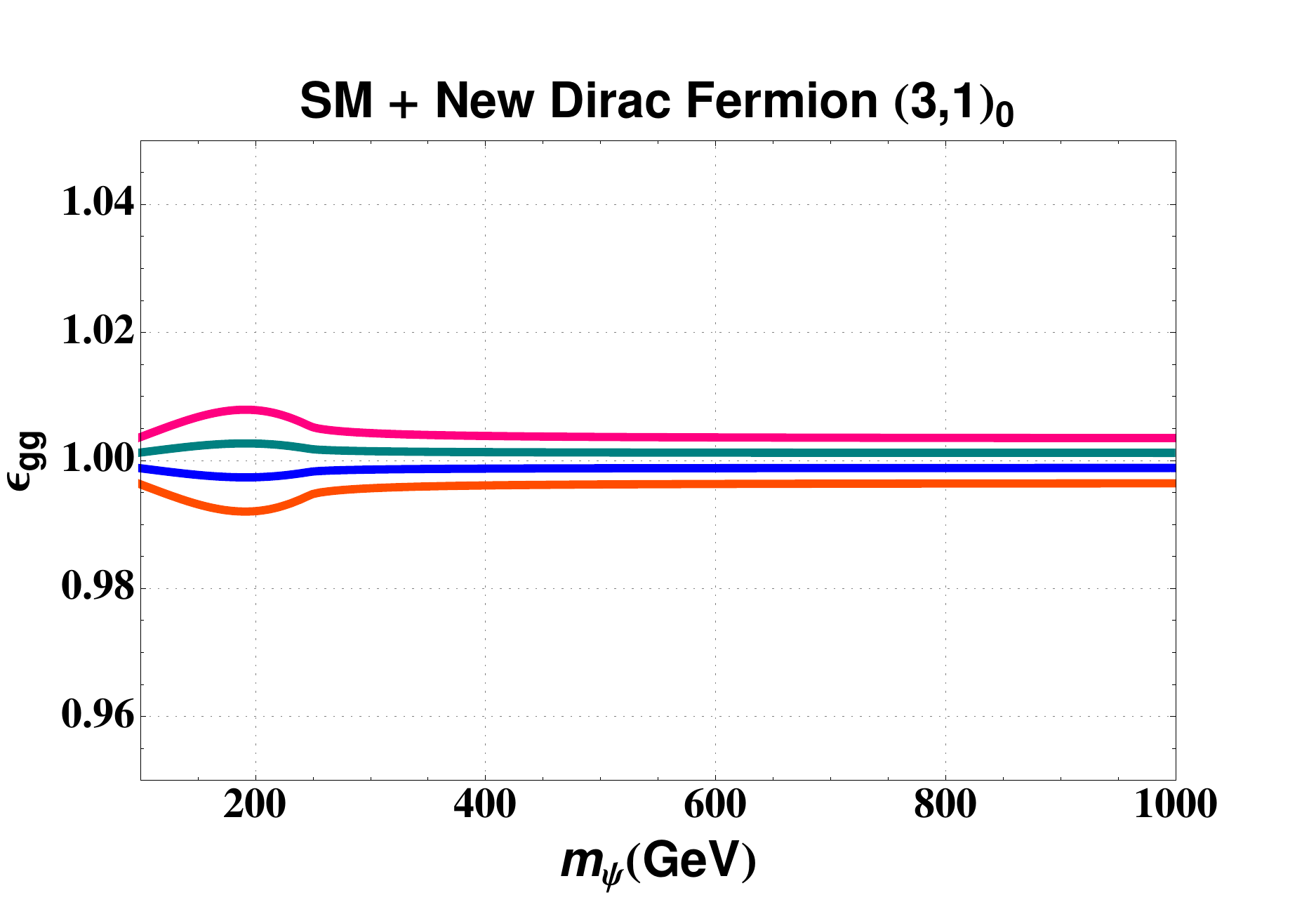}   
\caption{\label{fig:unmixed_fermion_relative_rate}{Relative rate of
    $s_1$ production with Higgs mixing and a new color octet fermion
    (left) and a new color triplet fermion (right).  We choose
    $v_{\phi}=500$ GeV, $m_{s_1}=500$ GeV and $m_{s_2} = 700$ GeV.
    From top to bottom, the solid lines in each plot correspond to
    $\lambda_{hp}=0.015$, $0.005$, $-0.005$, $-0.015$. }}
\end{center}
\end{figure}

Separately, we can also consider new colored fermions which do not mix
with the SM quarks.  As a final example, we consider new electroweak
singlet fermions $\psi$ in the adjoint and fundamental representations
of $SU(3)_c$ which obtain mass from the new Yukawa term
of~\eqnref{vectormass_generation} (with $\chi \rightarrow \psi$).  In
particular, these fermions do not couple to $h$, and hence their
effects will be suppressed by the scalar mixing angle
in~\eqnref{mixingangle}.  The relative rate is now
\begin{equation}
\label{eqn:SM+Fadj_epsilon}
\left. \epsilon_{gg} \right|_{SM + \psi}  = \dfrac{ c^2_\theta \left| 
\sum\limits_{f} \left( \dfrac{C(r_f)}{v_h} F_F(\tau_f) \right)
- t_\theta 
\left( \dfrac{C(r_\psi)}{v_\phi} F_F(\tau_\psi) \right)
\right|^2 }{ 
\left|
\sum\limits_{f} \left( \dfrac{C(r_f)}{v_h} F_F(\tau_f) \right)
\right|^2 } \ .
\end{equation}
We can see that $\psi$ does not decouple from the $gg \rightarrow s_1$
amplitude as its mass is taken very large because $F_F (\tau_\psi)$
asymptotes to a finite value.  We show $\epsilon_{gg}$
in~\figref{unmixed_fermion_relative_rate} for two choices of color
representations.  We see from~\figref{unmixed_fermion_relative_rate}
that the octet fermion (which includes a 1/2 to account for lack of
conjugate diagram for a real fermion) produces larger suppression or
enhancement than the triplet fermion for identical $\lambda_{hp}$
values, as expected from the difference in their respective
$C(r_{\psi})$.  Collider constraints on these new fermions will
require model dependent assumptions about their decay channels, and
since our focus is on the model independent effects on gluon fusion,
we do not consider such constraints here.

\section{New Colored Vector}
\label{sec:Vector}

The last type of New Physics contribution to gluon fusion we will
consider is the addtion of a new massive colored vector.  In a
renormalizable theory, the massive vector must arise from a
spontaneously broken gauge theory, which necessarily entails the
addition of a new scalar that acquires a vev and can mix with the SM
Higgs via~\eqnref{Lhp}.  Correspondingly, we will consider an extended
gauge symmetry $SU(3)_1 \times SU(3)_2$, known as the renormalizable
coloron model (ReCoM)~\cite{Hill:2002ap, Bai:2010dj}.  In this model,
the complex scalar field $\Phi$ transforms as $(3, \overline{3})$ and
obtains a diagonal vev (when written as a matrix-valued field), which
breaks $SU(3)_1 \times SU(3)_2$ to the diagonal subgroup, which is
identified with the SM $SU(3)_c$ symmetry.  The $\Phi$ field has 18
degrees of freedom: 8 are ``eaten'' by the broken gauge generators to
make the massive color vector $G'$ known as the coloron, 8 become a
real scalar $SU(3)_c$ octet $G_H$, and the remaining 2 are the real
scalar $\phi_R$ and pseudoscalar $\phi_I$ color singlet fields.
Hence, in this construction and a consequence of the Higgs portal, the
addition of a massive color vector $G^\prime$ concomitantly includes a
new scalar octet and two new scalar singlets, of which $G_H$
necessarily affects gluon fusion and $\phi_R$ mixes with the SM Higgs.

\subsection{The Renormalizable Coloron Model}
\label{subsec:ReCoM}
We analyze the total scalar potential including the SM, the ReCoM, and
the Higgs portal.  Our analysis mirrors that found
in~\cite{Bai:2010dj}, except our addition of the Higgs portal operator
creates a link between the two scalar potentials $V(H)$ and $V(\Phi)$
and hence the two vevs must be solved for simultaneously.  The full scalar
potential is
\begin{equation}
\label{eqn:fullscalarpotential}
\begin{array}{ccl}
V_{tot} &=& V(\Phi) + V(H) + V_{hp} \\
&=& -m_\Phi^2 \text{ Tr}(\Phi^\dagger \Phi) - \mu_\Phi(\det \Phi + 
\text{ h.c.})
+ \dfrac{\lambda_\Phi}{2} \left[ \text{Tr}( \Phi \Phi^\dagger ) \right]^2
+ \dfrac{\kappa_\Phi}{2} \text{ Tr}( \Phi \Phi^\dagger \Phi \Phi^\dagger) \\
&-& m_H^2 |H|^2 + \lambda_H |H|^4 \\
&-& \lambda_{hp} |H|^2 \text{ Tr}(\Phi^\dagger \Phi) \ , \\
\end{array}
\end{equation}
where we assume $\mu_\Phi > 0$ without loss of generality.  We require
$m_\Phi^2 > 0$ and $m_H^2 > 0$ such that $\Phi$ and $H$ will acquire
vevs.  We also require $3 \lambda_\Phi + \kappa_\Phi > 0$ and
$\lambda_H > 0$ so each individual potential is bounded from below: we
neglect renormalization group effects when discussing bounds on
potential parameters.

It is straightforward to find the vevs for $\Phi$ and $H$ by
decoupling the two equation system.  We find, in analogy
with~\cite{Bai:2010dj},
\begin{equation}
\label{eqn:vPhi}
\langle \Phi \rangle = \dfrac{v_\phi}{\sqrt{6}} \mathbb{I}_3 
= \dfrac{ \mu_\Phi + 
    \sqrt{\mu_\Phi^2 + \left( 2 (3\lambda_\Phi+\kappa_\Phi) 
                 - \dfrac{3 \lambda_{hp}^2}{\lambda_H} \right)
                 \left( 2 m_\Phi^2 + 
                  \dfrac{\lambda_{hp} m_H^2}{\lambda_H} \right) }}{
  \left( 2 (3\lambda_\Phi + \kappa_\Phi) -
    \dfrac{ 3\lambda_{hp}^2}{\lambda_H} \right)}
  \mathbb{I}_3 \ .
\end{equation}
If $\lambda_{hp}$ is too large, then it can generate a new ground
state at large field values of $h$ and $\phi$.  The resulting upper
bound on $\lambda_{hp}$ is $\lambda_{hp}^2 < \dfrac{2}{3} \lambda_H (3
\lambda_{\Phi} + \kappa_{\Phi})$, which we satisfy by requiring
$v_{\phi} > 0$.  Given~\eqnref{vPhi}, the Higgs vev is most easily
written as
\begin{equation}
\label{eqn:vH}
\langle H \rangle = \dfrac{1}{\sqrt{2}} 
\left( \begin{array}{c}
0 \\ v_h \\ 
\end{array} \right) = 
\dfrac{1}{\sqrt{2}}
\left( \begin{array}{c}
0 \\ 
\sqrt{ \dfrac{m_H^2}{\lambda_H} + \dfrac{\lambda_{hp} v_\phi^2}{2
    \lambda_H}} \\
\end{array} \right) \ ,
\end{equation}
and $v_h$ is fixed to be 246 GeV.

Expanding $\Phi$ around its vev, we get
\begin{equation}
\label{eqn:Phi}
\Phi = \dfrac{1}{\sqrt{6}} 
\left( v_\phi + \phi_R + i \phi_I \right) \mathbb{I}_3 
+ \left( G_H^a + iG_G^a \right) T^a \ ,
\end{equation}
where $\phi_R$ and $\phi_I$ are $SU(3)_c$ singlets and $G_H$ and $G_G$
are $SU(3)_c$ octets~\cite{Bai:2010dj}.  The $G_G$ comprise the
Goldstone bosons which give mass to the coloron, $G^\prime$.  The
Higgs is decomposed in the usual way,
\begin{equation}
\label{eqn:Higgs}
H = \dfrac{1}{\sqrt{2}} 
\left(
\begin{array}{c}
G^{\pm} \\ 
v_h + h + iG_o \\
\end{array}
\right)
\end{equation}
where $G_o$ and $G^{\pm}$ are the Goldstone bosons eaten by the
electroweak gauge bosons.

After the spontaneous symmetry breaking $SU(3)_1\times
SU(3)_2\rightarrow SU(3)_c$ and EWSB, mixing is induced between the
singlets $\phi_R$ and $h$.  This leads to the mass squared matrix in
the $(h, \phi_R)$ interaction basis given
in~\eqnref{scalar_mass_matrix} but with $m_{\phi}^2 \rightarrow
m_{\phi_R}^2$ and
\begin{equation}
\label{eqn:mass_matrix_terms}
m_h^2 = 2\lambda_H v_h^2 \ , \quad
m_{\phi_R}^2 = \dfrac{v_\phi^2}{3}(3\lambda_\Phi + \kappa_\Phi) -
\dfrac{\mu_\Phi v_\phi}{\sqrt{6}} \ ,
\end{equation}
where $v_h$ and $v_\phi$ depend on $\lambda_{hp}$.  The assumption of
$v_{\phi} > 0$ and the conditions $0 \leq m_h^2 \leq m^2_{\phi_R}$
imply
\begin{equation}
\label{eqn:condition5}
\mu_\Phi < \sqrt{\dfrac{2}{3}}(3\lambda_\Phi + \kappa_\Phi) v_\phi \ .
\end{equation}
By our assumptions, the right hand side of~\eqnref{condition5} is
positive definite and thus bounds $\mu_\Phi$ from above.  Our analysis
follows exactly the same steps as~\subsecref{mixing}, giving the
following results:
\begin{equation}
\label{eqn:hphiR_mixing}
\begin{array}{ccc}
\tan 2\theta &=& 
\dfrac{-2 \lambda_{hp} v_h v_{\phi} }{m_{\phi_R}^2 - m_h^2} \ , \\
s_1 &=& h \cos \theta - \phi_R \sin \theta \ , \\
s_2 &=& h \sin \theta + \phi_R \sin \theta \ . \\
\end{array}
\end{equation}
For the physical masses of $s_1$ and $s_2$, we obtain
\begin{equation}
\label{eqn:ms1_ms2}
\begin{array}{ccl}
\vspace{4pt} 
m^2_{s_1} &=& \dfrac{1}{2}(m_h^2 + m_{\phi_R}^2) -
\dfrac{1}{2}\sqrt{(-m_h^2 + m_{\phi_R}^2)^2 + 4 \lambda_{hp}^2 v_h^2
  v_{\phi}^2} \ , \\
m^2_{s_2} &=& \dfrac{1}{2}(m_h^2 + m_{\phi_R}^2) +
\dfrac{1}{2}\sqrt{(-m_h^2 + m_{\phi_R}^2)^2 + 4 \lambda_{hp}^2 v_h^2
  v_{\phi}^2} \ . \\
\end{array}
\end{equation}
For the physical masses of the remaining scalars in the spectrum we
find,
\begin{equation}
\label{eqn:scalarmasses}
m_{\phi_I}^2 = \sqrt{\dfrac{3}{2}}\mu_\Phi v_\phi \ , \quad
m_{G_H}^2 = \dfrac{1}{3}(2m_{\phi_I}^2 + \kappa_\Phi v^2_\phi) \ ,
\end{equation}
which agrees with~\cite{Bai:2010dj} in the limit $\lambda_{hp}
\rightarrow 0$.

The constraint $m_{\phi_I}^2 > 0$ is satisfied since we assumed
$\mu_\Phi > 0$ and $v_\phi > 0$.  Requiring $m_{G_H}^2 > 0$ implies
$m_{\phi_I}^2 > -\kappa_\Phi v_\phi^2 /2$, which
augments the previous condition~\eqnref{condition5} to give
\begin{equation}
\label{eqn:mu_bounds}
-\frac{\kappa_\Phi v_\Phi}{\sqrt{6}} < \mu_\Phi <
\sqrt{\frac{2}{3}} \left( 3 \lambda_\Phi + \kappa_\Phi \right) v_\phi
\end{equation}
In order for a valid range of $\mu_\Phi$ to exist, we thus require
\begin{equation}
\label{eqn:lambda_kappa_inequality}
2 \lambda_\Phi + \kappa_\Phi > 0 \ .
\end{equation}
Our subsequent analysis ensures these constraints are satisfied.

\subsection{Phenomenology}
\label{subsec:SM3pVectorPheno}

\begin{figure}[htb]
\includegraphics[scale=0.6, angle=0]{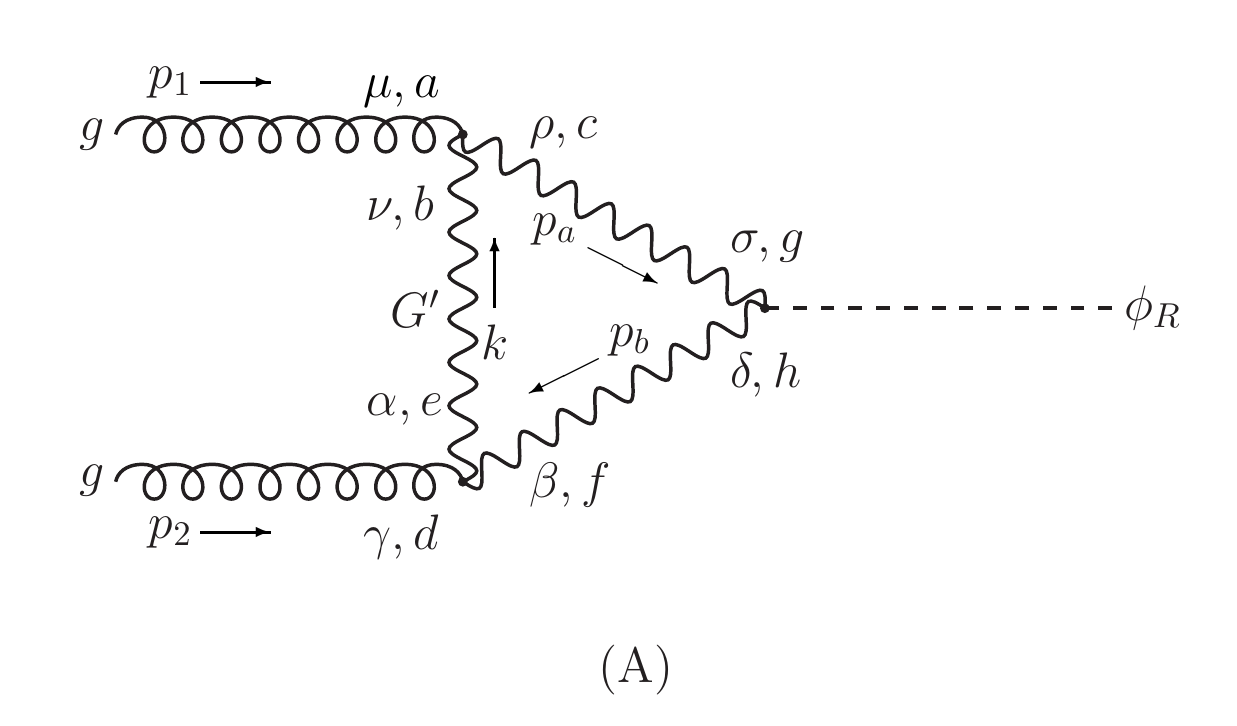}
\includegraphics[scale=0.6, angle=0]{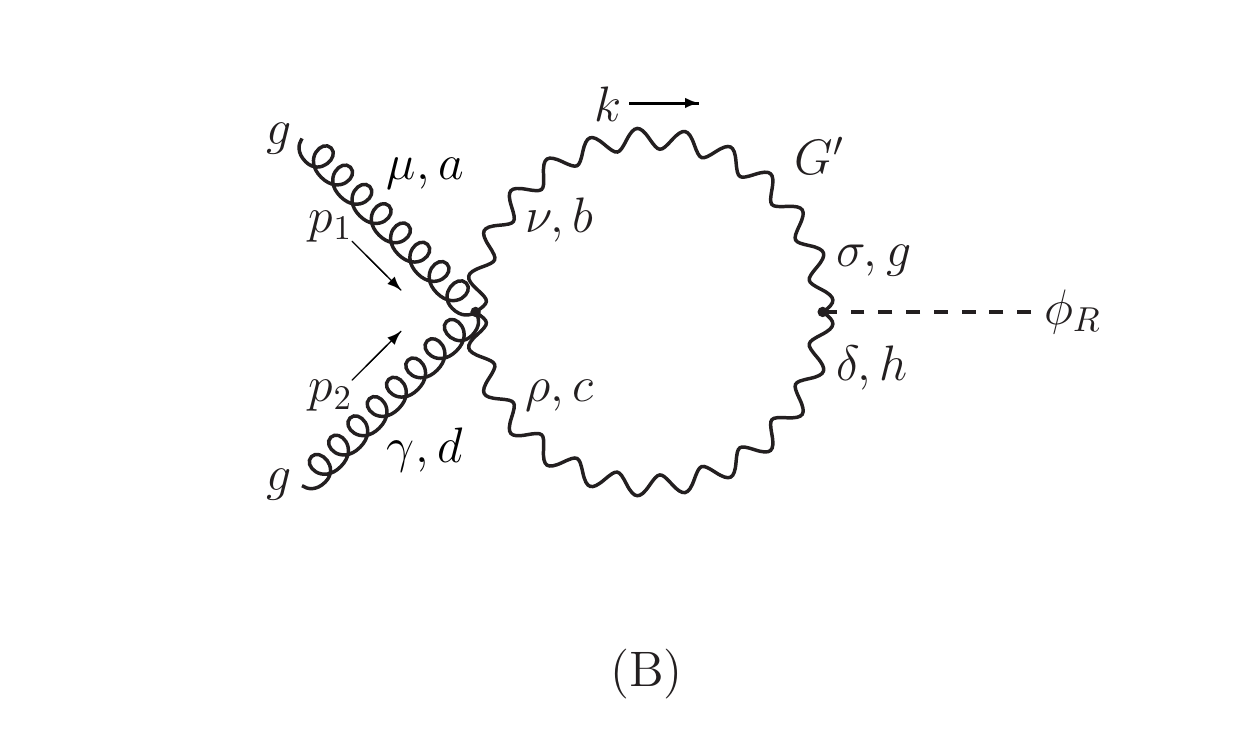}
\caption{\label{fig:Unitaryvectorloop}{Feynman diagrams for vector
    loop contributions to $gg \rightarrow \phi_R$ in the unitary
    gauge.}}
\end{figure}

The diagrams for the colored vector loop in unitary gauge are shown
in~\figref{Unitaryvectorloop}.  Although the ReCoM model includes the
$G_H$ scalar octet contribution, we have isolated colored scalar
amplitudes in~\secref{Scalar}, and so we focus here on the colored
vector contribution.

\begin{figure}[htb]
\includegraphics[width=0.7\textwidth, angle=0]{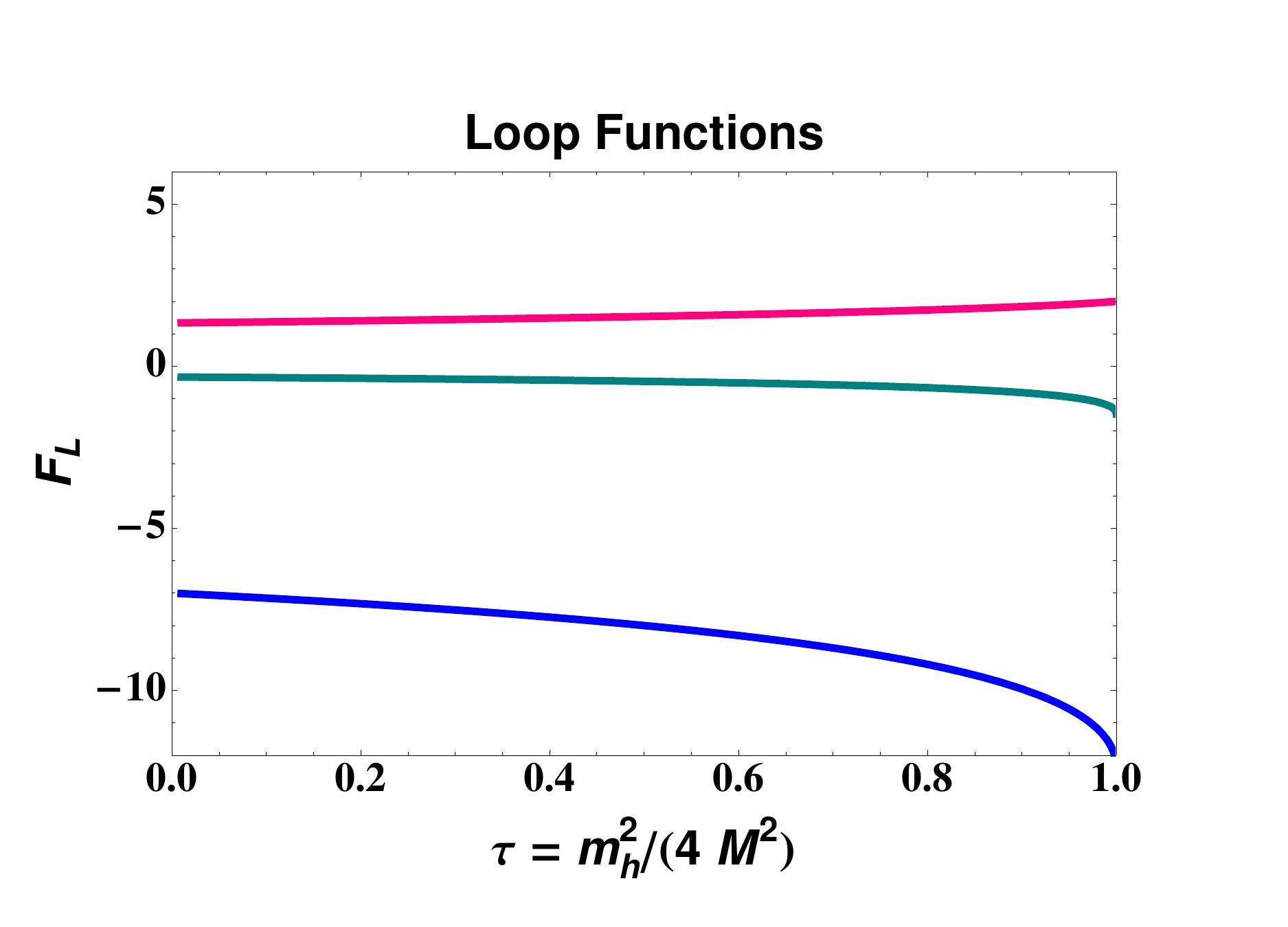}
\caption{\label{fig:loop_functions}{Loop functions $F_L = F_F$ (top),
    $F_L = F_S$ (middle), and $F_L = F_V$ (bottom) for fermion,
    scalar, and vector particles of mass $M$, respectively.  The loop
    functions develop an imaginary part for $\tau > 1$, which
    corresponds to the particles in the loop going on-shell. }}
\end{figure}

The full amplitude for a real vector field propagating in the loop is
\begin{equation}
\label{eqn:UGampsum}
\begin{array}{ccl}
i \mathcal{M}^{ad}_V &=& i \left. \mathcal{M}^{ad}_V 
(gg \rightarrow \phi_R) \right|_{m_{\phi_R} = m_{s_1}} \\
&=& i \left( \dfrac{\alpha_s}{\pi} \right)
\left( \dfrac{C(r_{G^\prime})}{4 v_\phi} \right) 
\delta^{ad} \epsilon_{1\mu} \epsilon_{2\gamma} 
\left( p_1^\gamma p_2^\mu - \dfrac{m_{s_1}^2}{2} g^{\mu\gamma} \right)
F_V(\tau_{G^\prime}) \\
\end{array}
\end{equation}
where
\begin{equation}
\label{eqn:FV_definition}
F_V(\tau) \equiv 
- \left( \tau^{-1} (3 + 2 \tau) 
+ 3 \tau^{-2} (-1 + 2 \tau)Z(\tau) \right) \ .
\end{equation}
A full derivation of this amplitude in both unitary and Feynman gauge
is presented in~\appref{Vector_explicit}.  The resulting enhancement factor is
\begin{equation}
\label{eqn:SM+V_epsilon}
\begin{array}{l}
\left. \epsilon_{gg} \right|_{SM + ReCoM} = \\
\dfrac{ c^2_\theta \left| 
\sum\limits_{f} \left( \dfrac{C(r_{f})}{2v_h} F_F(\tau_{f}) \right)
+  \dfrac{1}{2}  \dfrac{C(r_{G_H})}{4v_\phi} 
\left( \dfrac{ \lambda_{hp} v_h v_\phi - t_\theta x_{G_H} }{m_{G_{H}}^2} 
\right) F_S (\tau_{G_H})
- t_\theta 
\left(\dfrac{C(r_{G^\prime})}{4v_\phi} F_V(\tau_{G^\prime}) 
 \right)
\right|^2 }{ 
\left|
\sum\limits_{f} \left( \dfrac{C(r_f)}{2v_h} F_F(\tau_f) \right)
\right|^2 } \ , \\
\end{array}
\end{equation}
where $x_{G_H} / v_\phi = \left( -m_{\phi_R}^2 + \frac{2}{3}
m_{\phi_I}^2 - 2 m_{G_H}^2 \right) / v_\phi$ evaluated at $m_{\phi_R}
= m_{s_1}$ in~\eqnref{SM+V_epsilon} is the $G_H$ coupling to $\phi_R$
and we have included both $G_H$ (with an explicit symmetry factor of
1/2) and $G^\prime$ contributions. We note that the real colored
vector loop function $F_{G^\prime}$ is numerically about a factor of 5
larger and of the opposite sign than the usual SM loop function $F_F$.
The scalar loop function $F_S(\tau)$ is also of the opposite sign and
roughly a third of $F_F$: the loop functions are plotted
in~\figref{loop_functions}.  We comment that, as a result of the large
loop function for the colored vector, moderate values of
$\lambda_{hp}$ can have large effects on gluon fusion production.
This provides a straightforward construction, for example, to
counteract the enhancement from a fourth generation of chiral fermions
in the Standard Model.  If such a large cancellation of $gg
\rightarrow h$ amplitudes was present, however, we expect di-Higgs
production via $gg \rightarrow hh$ to be correspondingly enhanced if
the $gg \rightarrow hh$ triangle and bubble amplitudes are negligible:
this is because the individual signs of direct Higgs couplings that
lead to suppression become squared in the $gg \rightarrow hh$ box
amplitude.  In this case, the di-Higgs gluon fusion production channel
may be a promising discovery mode, and a more careful study is
certainly warranted.

\begin{figure}[htb]
\includegraphics[width=0.48\textwidth]{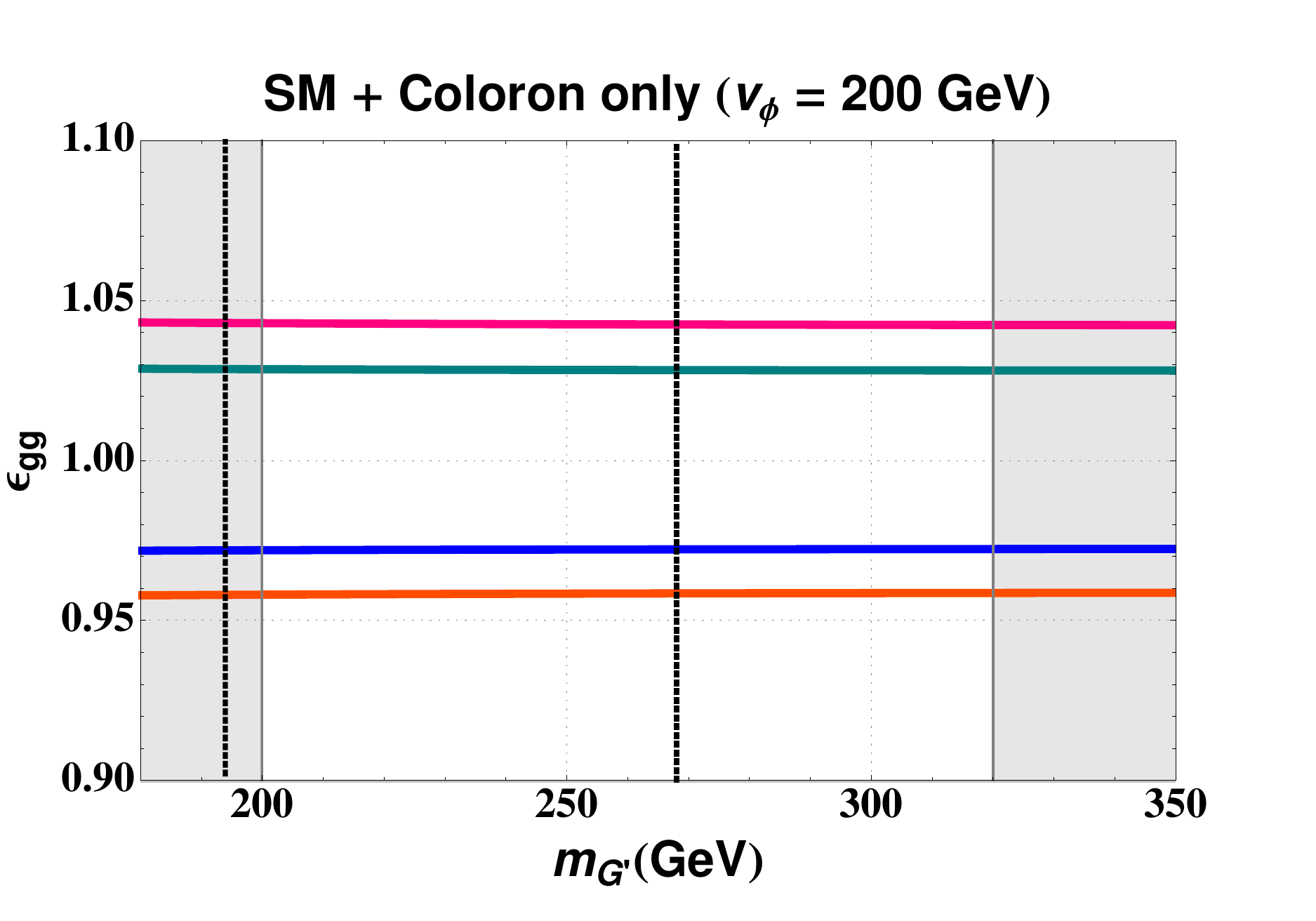}
\includegraphics[width=0.48\textwidth]{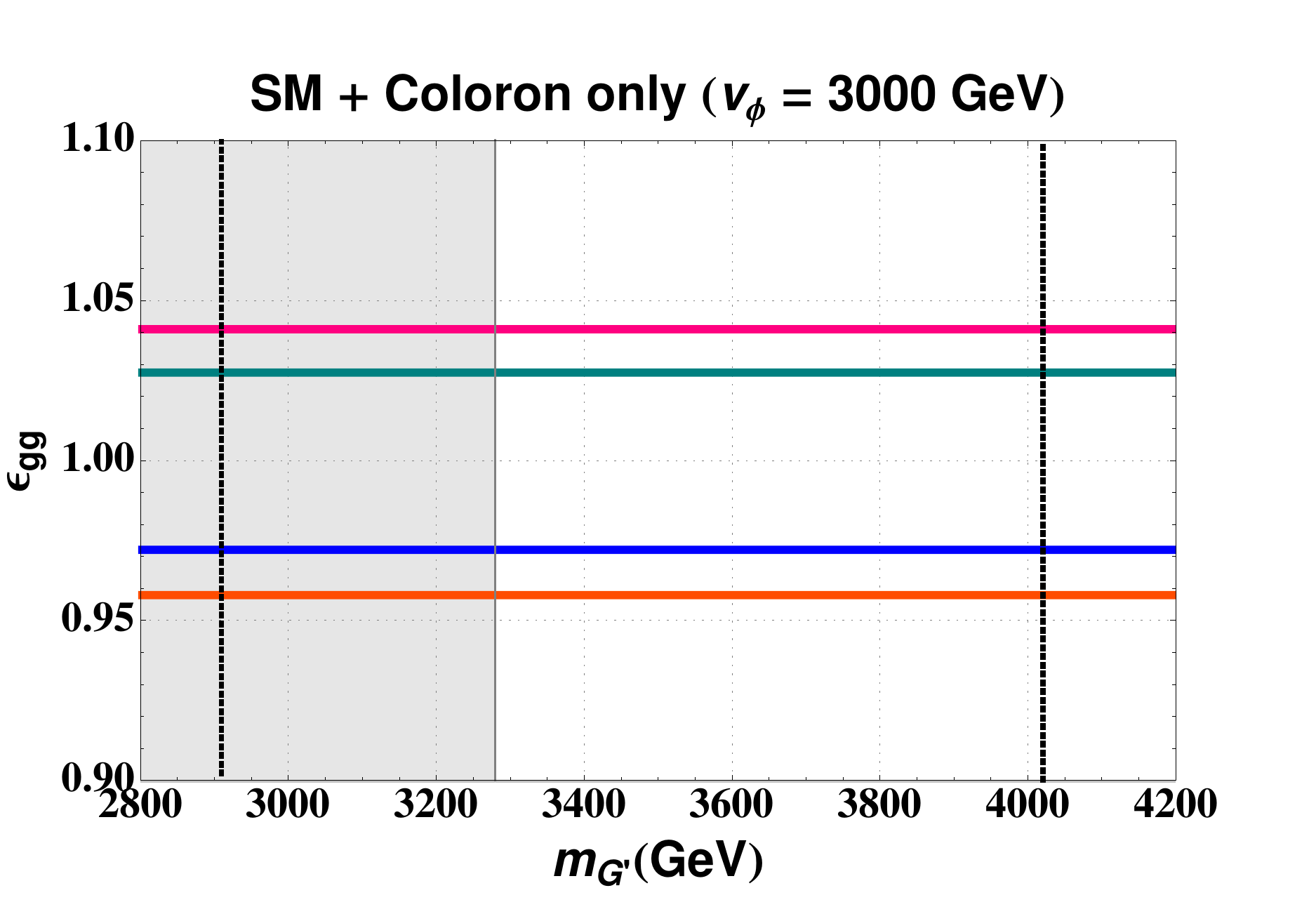}
\includegraphics[width=0.48\textwidth, angle=0]{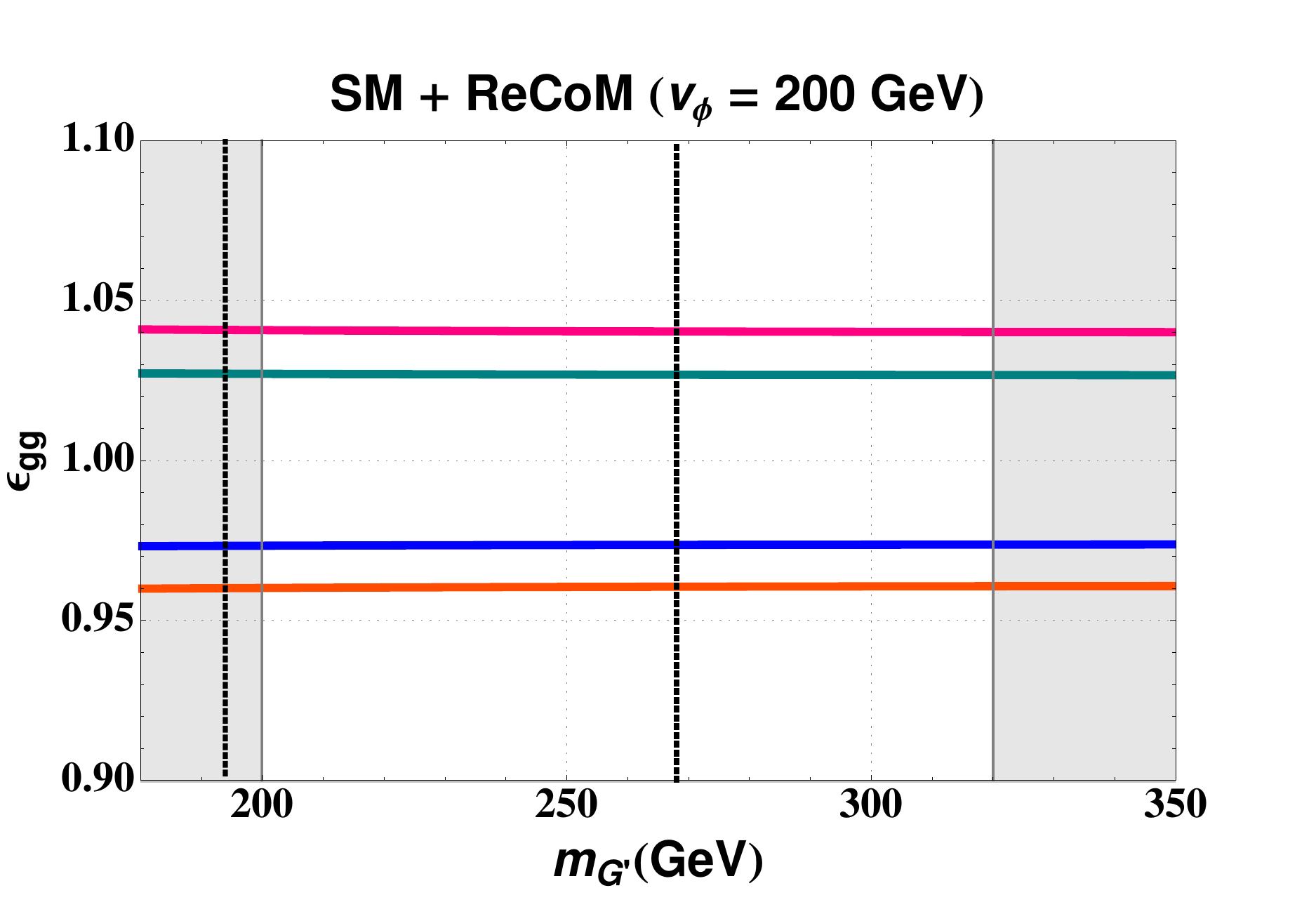}
\includegraphics[width=0.48\textwidth, angle=0]{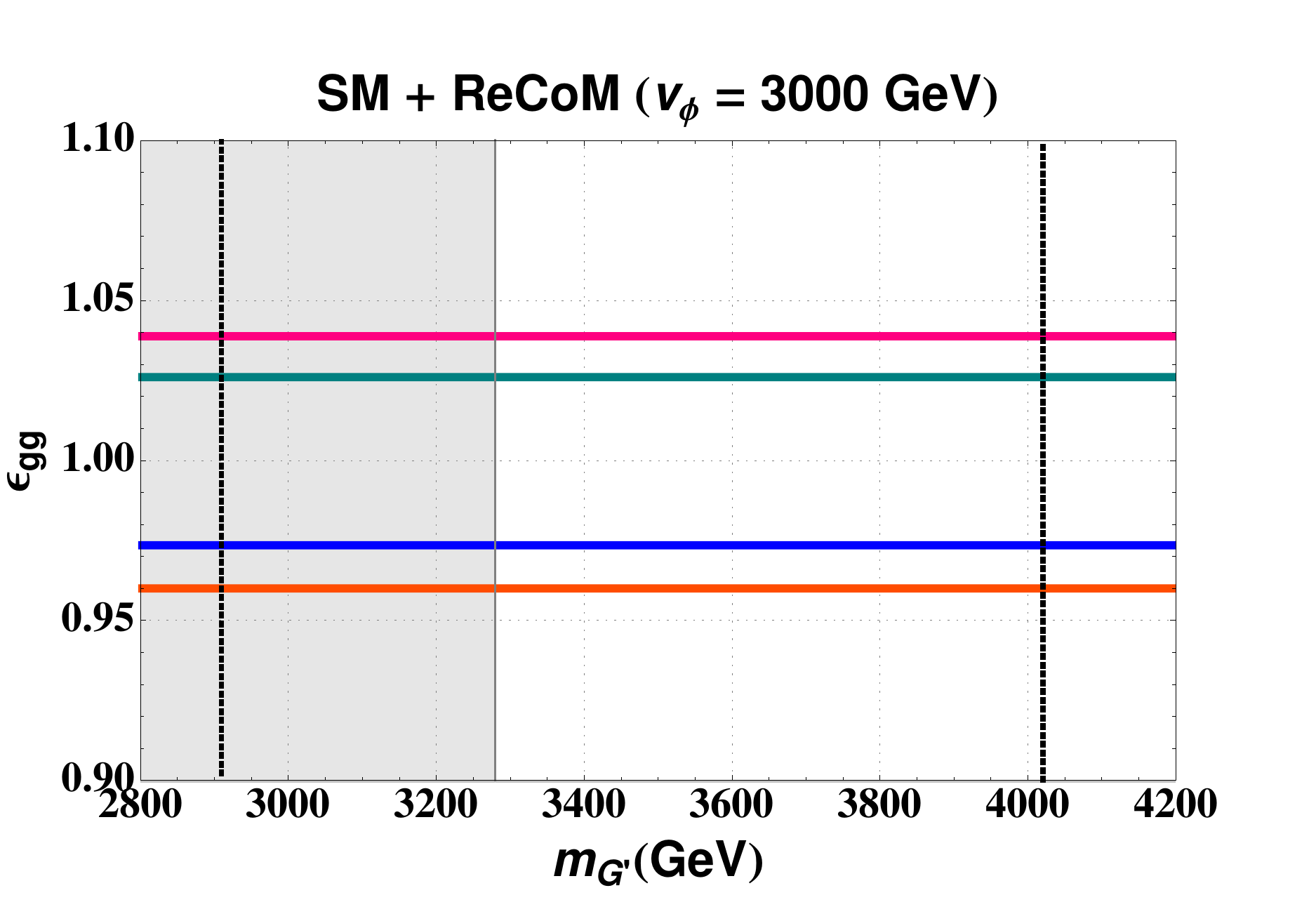}
\caption{\label{fig:vector_LMsContour_fixmh}{Relative rate of $s_1$
    production showing the effect of adding only the coloron (top row)
    and the effect of adding the coloron and $G_H$, with $m_{G_H} =
    140$ GeV (bottom row).  From top to bottom, the solid lines
    correspond to $\lambda_{hp} = -7.5 \times 10^{-4}$, $-5 \times
    10^{-4}$, $5 \times 10^{-4}$, $7.5 \times 10^{-4}$. In the left
    panels, the gray bands correspond to the pair-produced dijet
    bounds from ATLAS (left) and CMS (right).  In the right panels,
    the left gray band corresponds to the CMS exclusion on coloron
    production in dijet resonance searches.  The left vertical dotted
    line corresponds to the minimum $m_{G'}$ mass allowed given the
    specified choice of $v_\phi$, and the right vertical dotted line
    corresponds to the perturbativity constraint. }}
\end{figure}

We first present results in~\figref{vector_LMsContour_fixmh} for
$\epsilon_{gg}$ with the sole addition of the colored vector for
$v_\phi = 200$ GeV and $3$ TeV.  The $v_\phi = 200$ GeV choice is
interesting because bounds on colorons in the low mass region coming
from pairs of dijet resonances from ATLAS~\cite{Aad:2011yh} and
CMS~\cite{CMS-PAS-EXO-11-016} are not currently sensitive in the 200
to 320 GeV range, as discussed in~\secref{Scalar}. There are numerous
recent studies which have done fits of the Higgs couplings to the
data, including the Higgs-gluon effective coupling~\cite{Carmi:2012yp,
  Carmi:2012zd, Giardino:2012ww, Giudice:2007fh}. For $v_{\phi}=200$
GeV, a coloron mass of $250$ GeV and mixing angle of $s_\theta \sim
-0.04$ one can reproduce the best fit point ($c_g \sim 0.5$) for the
scalar-gluon effective coupling found in~\cite{Carmi:2012zd} and as
can be seen from~\figref{vector_LMsContour_fixmh} one can also easily
reproduce values which give rates close to the SM value should the
effective coupling migrate towards the SM prediction with more
data. We also consider $v_\phi = 3$ TeV, which is the scale probed in
dijet resonance searches using 4 fb$^{-1}$ of 8 TeV LHC data at
CMS~\cite{CMS-PAS-EXO-12-016}. The observed limit from this search on
the coloron mass, $m_{G^\prime}$, is 3.28 TeV. The bounds are
indicated by gray vertical bands as described in the caption. The
latest dijet resonance search done by the ATLAS
collaboration~\cite{ATLAS-CONF-2012-088} with 5.8 fb$^{-1}$ of 8 TeV
LHC data does not report an observed limit on the coloron mass but we
expect the limit to be just a little higher because of the increased
luminosity. An additional constraint on the coloron mass arises from
the constraints imposed by ReCoM, whereby perturbativity restrictions
on the original $SU(3)_1 \times SU(3)_2$ gauge couplings give an upper
limit and requirements on generating the correct $SU(3)_c$ coupling
give a lower limit.  In deriving these bounds, which are given by
dashed vertical lines in~\figref{vector_LMsContour_fixmh}, we have
neglected renormalization group running of $\alpha_s$.

In general, the ReCoM model includes contributions from the color
vector $G^\prime$ and the scalar octet $G_H$.  We can see
from~\eqnref{SM+V_epsilon} that the contribution from $G_H$ coming
through the $h$ component of $s_1$ always adds constructively with the
$G^\prime$ contribution.  Whether the contribution from $G_H$ entering
through $\phi_R$ also adds constructively with the $G^\prime$
contribution depends on the sign of $x_{G_H}$, which in turn depends
on the mass hierarchy between the various scalars.  We present the
effect arising from only the color vector in the top row and from both
new colored states in the bottom row
of~\figref{vector_LMsContour_fixmh}.  For these plots, we have set
$m_{s_1} = 125$ GeV, $m_{s_2} = 225$ GeV, $m_{\phi_I} = 160$ GeV, and
$m_{G_H} = 140$ GeV.\footnote{The ATLAS search for dijet pairs has an
  upward fluctuation above their expected limit in the 140 GeV range,
  which leaves the window of a scalar octet open for this mass point.}

We remark that the flat behavior of $\epsilon_{gg}$ in each plot
arises because for $m_{G^\prime} > m_{s_1}$, the loop function
dependence of $m_{G^\prime}$ asymptotes quickly. This reflects the
fact that as $m_{G^\prime}$ is taken large, its effects (which enter
only through the $\phi_R$ component of $s_1$) do not decouple from the
$s_1$ production amplitude, which is reminiscent of the non-decoupling
of $W$ bosons from the SM Higgs in $h \rightarrow \gamma \gamma$.  In
addition, for the scalar octet $G_H$, which couples to both $h$ and
$\phi_R$ components, we find that as $m_{G_H}$ is taken large, its
effects do not decouple from the $\phi_R$ component but do decouple
from $h$, see~\secref{Scalar}.  Finally, we note the small reduction
of $\epsilon_{gg}$ in ReCoM is a result of the $G_H$ contribution
slightly cancelling the $G^\prime$ contribution, given our chosen
parameter point for which $x_{G_H} < 0$, and we see that the overall
effect is dominated by the coloron contribution, as expected from the
magnitudes of the loop functions shown in~\figref{loop_functions}.

\section{Discussion and Conclusion}
\label{sec:Conclusion}

We have seen that new colored particles can have significant effects
on gluon fusion production.  Also, the inclusion of Higgs
portal-induced scalar mixing readily leads to new possibilities for
suppressing or enhancing the gluon fusion rate.  We have isolated
contributions arising from new colored scalars, new colored fermions,
including quark mixing, and new colored vectors.  With the amplitudes
in~\eqnref{scalar_ME},~\eqnref{exotic_fermionamp},
and~\eqnref{UGampsum}, we can immediately calculate the interference
effects present in general new physics models.  We have demonstrated
these effects can easily run from $\mathcal{O}(1\%)$ to
$\mathcal{O}(10\%)$, and both suppression and enhancement of gluon
fusion can occur.  Moreover, such large deviations are possible by
colored states at mass scales that can be directly probed at the LHC.
In particular, when the effects on gluon fusion scale with the Higgs
portal coupling $\lambda_{hp}$, the dearth of restrictions on
$\lambda_{hp}$ from the underlying theory can lead to very large
modifications.  If many competing effects are present in the $gg
\rightarrow h$ amplitude, we expect that di-Higgs production will be
correspondingly altered and a promising discovery mode, but we leave a
more careful study for future work.  

Since gluon fusion production does not directly probe the Higgs
mechanism or the phenomenon of electroweak symmetry breaking, Higgs
identification studies should generally allow for mixing with a
separate scalar state as well as competing effects from multiple new
colored states running in the loop.  Our general framework and analysis can be easily mapped onto the various recent studies which attempt to fit the data with non-SM Higgs couplings to two gluons. In particular we have shown that the various new physics effects can conspire to give rates very close to the SM expectation as well as easily accounting for any slight deviations as suggested by recent fits of the scalar effective couplings~\cite{Carmi:2012yp, Carmi:2012zd, Giardino:2012ww, Giudice:2007fh}.

\section*{Acknowledgements}
\label{sec:acknowledgements}
The authors are grateful to Bill Bardeen, Marcela Carena, Andr\'{e} de
Gouv\^{e}a, Bogdan Dobrescu, Jennifer Kile, Ian Low, Adam Martin,
Reinard Primulando, Pedro Schwaller, Daniel Stolarski, and Tim Tait
for useful discussions.  Fermilab is operated by Fermi Research
Alliance, LLC under Contract No. De-AC02-07CH11359 with the United
States Department of Energy.

\begin{appendix}

\section{Vector Loop Calculation}
\label{sec:Vector_explicit}

In this appendix, we present the explicit calculation of the vector
loop contribution to gluon fusion.  As the loop calculation involves a
particular choice of $R_\xi$ gauge, we perform the calculation in both
unitary ($R_{\xi} \rightarrow \infty$) and Feynman ($R_{\xi} = 1$)
gauges.  This calculation generalizes the well-known Standard Model
calculation for $h \rightarrow \gamma \gamma$~\cite{Ellis:1975ap,
  Shifman:1979eb, Marciano:2011gm} to situations where a new ``Higgs''
field acquires a vev and leaves a non-Abelian gauge symmetry unbroken.
In the SM, the Higgs field is responsible for spontaneously breaking
$SU(2)_L \times U(1)_Y$, leaving the photon as the gauge field of the
remaining Abelian $U(1)_{em}$ gauge symmetry.  In contrast, in the
renormalizable coloron model (ReCoM), $\Phi$ is responsible for
spontaneously breaking $SU(3)_1 \times SU(3)_2$, leaving the gluon as
the gauge field of the remaining non-Abelian $SU(3)_c$ gauge symmetry.
Then, $gg \rightarrow \phi_R$ is the non-Abelian mirror version of $h
\rightarrow \gamma \gamma$.  We intuit that $\mathcal{M}(gg
\rightarrow \phi_R)$ is enhanced by a color factor over the mirror
process $\mathcal{M}(h \rightarrow \gamma \gamma)$, which is borne out
from our calculation.

\subsection{Vector Loop Amplitude: Unitary Gauge}
\label{subsec:Vector_unitary_explicit}
We present the unitary gauge calculation of a colored vector
contribution to gluon fusion.  As mentioned above, we assume an
extended color gauge symmetry that is partially broken by the vev of a
new scalar field $\Phi$.  After Higgs portal-induced mixing of $h$ and
$\phi_R$, the matrix elements $\mathcal{M}(gg \rightarrow s_1)$ and
$\mathcal{M}(gg \rightarrow s_2)$ are simply related to
$\mathcal{M}(gg \rightarrow h)$ and $\mathcal{M}(gg \rightarrow
\phi_R)$ by
\begin{equation}
\label{eqn:ME_projections}
\begin{array}{ccc}
\mathcal{M} (gg \rightarrow s_1) &=& 
\left. 
c_\theta \left[ \mathcal{M} (gg \rightarrow h) \right]
\right|_{m_h = m_{s_1}}
\left. 
-s_\theta \left[ \mathcal{M} (gg \rightarrow \phi_R) \right]
\right|_{m_{\phi_R} = m_{s_1}} \\
\mathcal{M} (gg \rightarrow s_2) &=& 
\left.
 s_\theta \left[ \mathcal{M} (gg \rightarrow h) \right]
\right|_{m_h = m_{s_2}}
\left.
+c_\theta \left[ \mathcal{M} (gg \rightarrow \phi_R) \right]
\right|_{m_{\phi_R} = m_{s_2}} \ . \\
\end{array}
\end{equation}

There are two diagrams which contribute to $\mathcal{M} (gg \rightarrow
\phi_R)$ in the unitary gauge, shown in~\figref{Unitaryvectorloop} of
the main text.  The triangle diagram in~\figref{Unitaryvectorloop}A
for the coloron insertion gives
\begin{eqnarray}
\label{eqn:MaAmplitude}
i{\mathcal M}^{ad}_A &=&
-g_s^2 \left( \dfrac{2m_{G^\prime}^2}{v_\phi}\right) f^{abc}f^{dcb}  
\epsilon_{1\mu} \epsilon_{2\gamma} \nonumber \\
& & \int \dfrac{d^dk}{(2\pi)^d}
\dfrac{V^{\mu\nu\rho} 
\left( g_{\alpha \nu} - \dfrac{k_\alpha k_\nu}{m_{G^\prime}^2}\right) 
V^{\gamma\beta\alpha}
\left(g_{\beta \delta} - \dfrac{{p_b}_\beta {p_b}_\delta}{m_{G^\prime}^2}\right) 
g^{\sigma\delta}
\left(g_{\sigma \rho} -  \dfrac{{p_a}_\sigma {p_a}_\rho}{m_{G^\prime}^2}\right)
}{(k^2 - m_{G^\prime}^2)(p_a^2 - m_{G^\prime}^2)(p_b^2 - m_{G^\prime}^2)} \ ,
\end{eqnarray}
where $p_a = k + p_1$, $p_b = k - p_2$, and the three-point vector vertex
is
\begin{equation}
\begin{array}{ccc}
\label{eqn:three_vertex}
V^{\mu\nu\rho} = (k + p_a)^\mu g^{\nu \rho} + (-p_a - p_1)^\nu g^{\rho
  \mu} +(p_1 -k)^\rho g^{\mu \nu} \\
V^{\gamma\beta\alpha} = (p_b + k)^\gamma g^{\alpha \beta} + (p_2 -
p_b)^\alpha g^{\gamma \beta} +(-k - p_2)^\beta g^{\alpha \gamma} \ . \\
\end{array}
\end{equation}
The amplitude for the bubble loop in~\figref{Unitaryvectorloop}B is
\begin{equation}
\label{eqn:MbAmplitude}
i{\mathcal M}^{ad}_B =
- \left( \dfrac{1}{2} \right) 
g_s^2 \left(\dfrac{2m_{G^\prime}^2}{v_\phi}\right) 
\epsilon_{1\mu} \epsilon_{2\gamma} \int \dfrac{d^dk}{(2\pi)^d}
\dfrac{V_{acdb}^{\mu\rho\gamma\beta} g^{\sigma \delta}
\left(g_{\delta \beta} - \dfrac{p_{a \delta} p_{a \beta}}{m_{G^\prime}^2}\right)
\left(g_{\rho \sigma} - \dfrac{k_\rho k_\sigma}{m_{G^\prime}^2}\right)}{(p_a^2 -
  m_{G^\prime}^2)(k^2 - m_{G^\prime}^2)} \ ,
\end{equation}
where $p_a = p_1 + p_2 - k$, a symmetry factor of $\dfrac{1}{2}$ has
been included, and the four-point vector vertex is
\begin{equation}
\label{eqn:four_vertex}
\begin{array}{ccl}
\delta^{bc} V_{acdb}^{\mu\rho\gamma\beta} &=& 
-ig_s^2 \delta^{bc} \left(f^{ace}f^{dbe}
\left(g^{\mu \gamma}g^{\rho \beta}-g^{\mu \beta}g^{\rho \gamma}\right) 
+ f^{ade}f^{cbe}
\left(g^{\mu \rho}g^{\gamma \beta}-g^{\mu \beta}g^{\gamma \rho}\right) \right. \\
&+& f^{abe}f^{cde} \left.
\left(g^{\mu \rho}g^{\gamma \beta} - g^{\mu \gamma}g^{\rho \beta}\right) \right) \\
&=&
-ig_s^2 f^{abe}f^{dbe}
\left( 2 g^{\mu \gamma} g^{\rho \beta} - g^{\mu \beta} g^{\rho \gamma}
 - g^{\mu \rho} g^{\gamma \beta} \right) \ .
\end{array}
\end{equation}
After expanding both~\eqnsref{MaAmplitude}{MbAmplitude} using Feynman
parameters, performing the loop momentum integration using Dimensional
Regularization~\cite{'tHooft:1972fi}, and some simpification, we
arrive at the summed amplitude
\begin{equation}
\label{eqn:Unitary_amplitude}
i{\mathcal M}^{ad}_V = i \left( \dfrac{\alpha_s}{\pi}  \right) 
\left( \dfrac{C(r_{G^\prime})}{4 v_\phi} \right) 
\delta^{ad} \epsilon_{1\mu} \epsilon_{2\gamma} 
\left( p_1^\gamma p_2^\mu - \dfrac{m_{s_1}^2}{2} g^{\mu\gamma} \right)
F_V(\tau_{G^\prime}) \ ,
\end{equation}
where $C(r_{G^\prime}) = 3$ for the coloron, $\tau_{G^\prime} =
m_{s_1}^2 / (4 m_{G^\prime}^2)$, and the loop function $F_V$ is given
in the main text in~\eqnref{FV_definition}.  We remark that this
result also agrees with the analogous SM calculation for $h
\xrightarrow[W]{} \gamma \gamma$ with the appropriate substitutions
$\alpha_s \rightarrow \alpha$, $C(r) \rightarrow 1$, $v_\phi
\rightarrow v_h$, and a factor of 2 included for the charge conjugate
process~\cite{Ellis:1975ap, Shifman:1979eb, Marciano:2011gm}.

\subsection{Calculation: Feynman Gauge}
\label{subsec:Vector_Feynman_explicit}

\begin{figure}[htb]
\includegraphics[scale=0.45, angle=0]{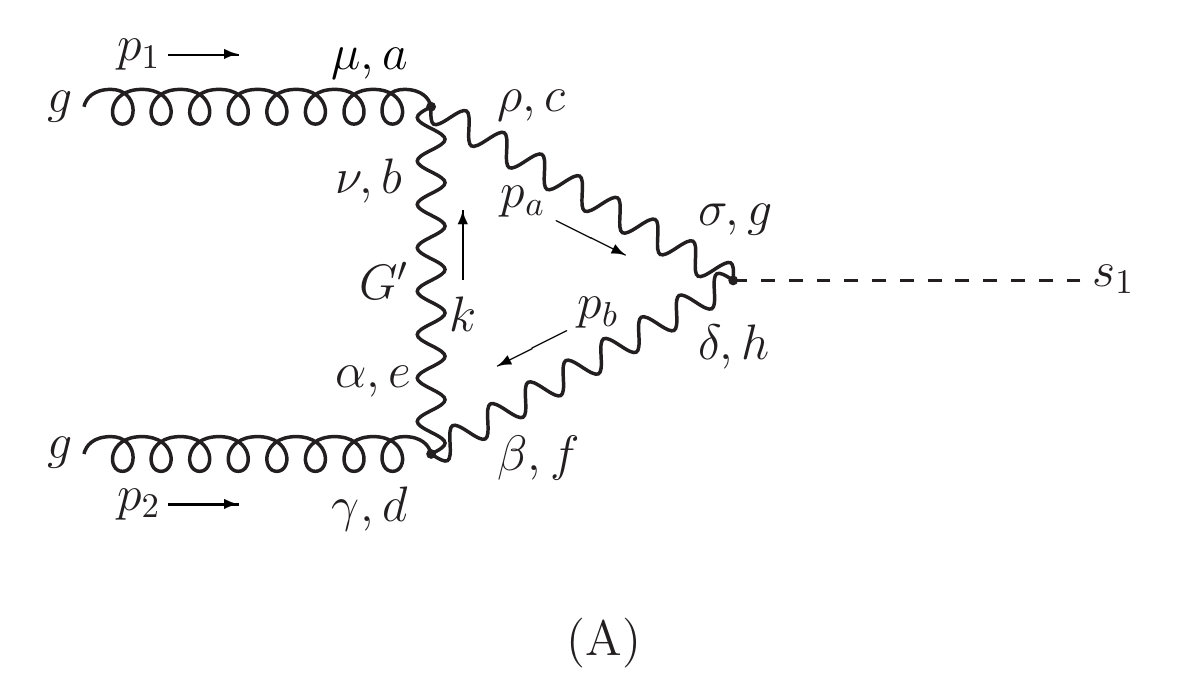}
\includegraphics[scale=0.45, angle=0]{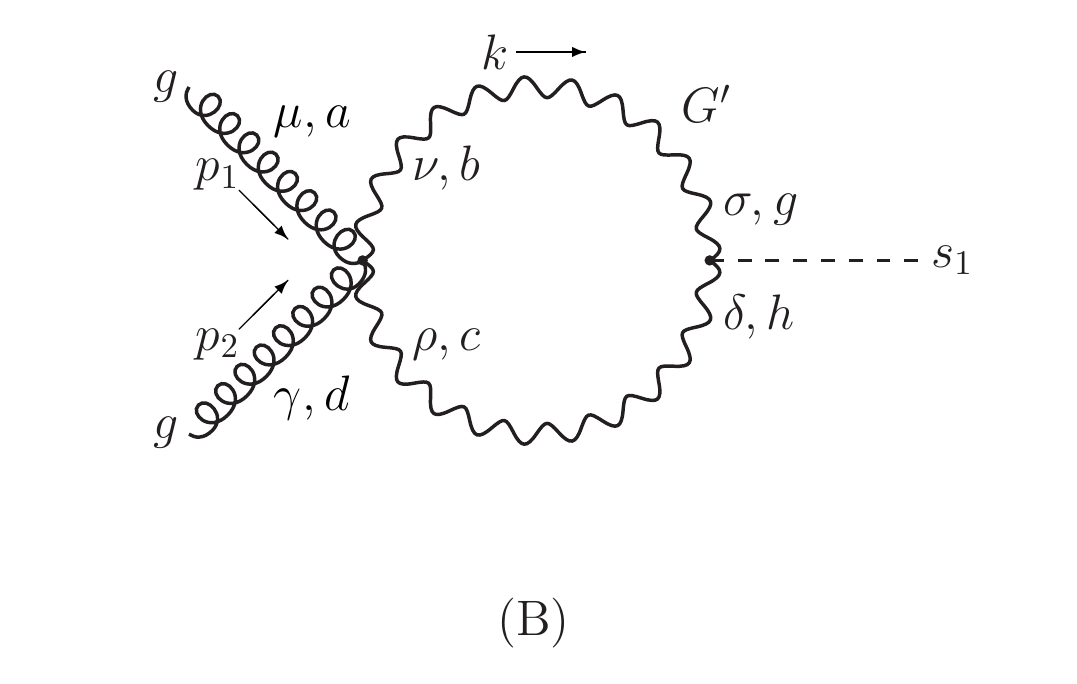}
\includegraphics[scale=0.45, angle=0]{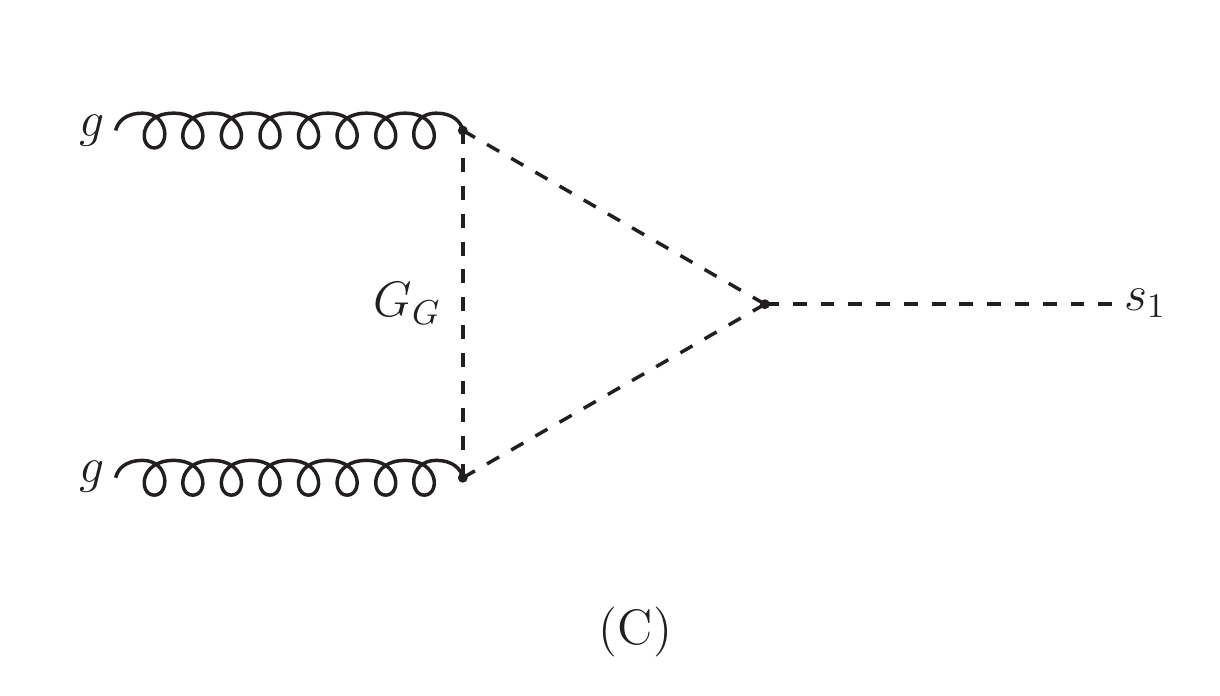}
\includegraphics[scale=0.45, angle=0]{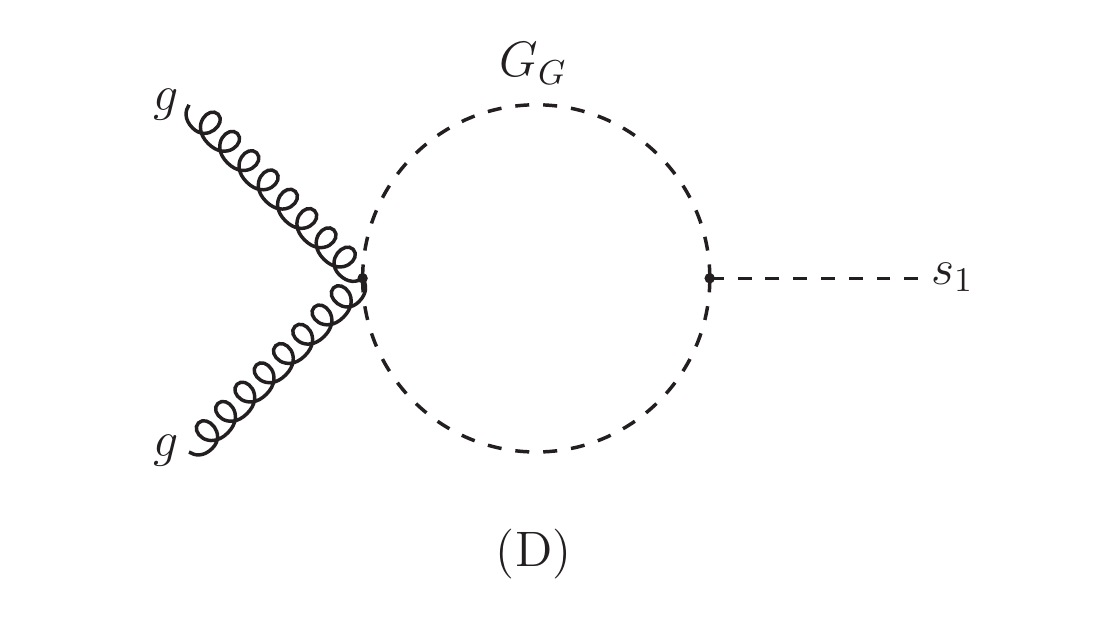}
\includegraphics[scale=0.45, angle=0]{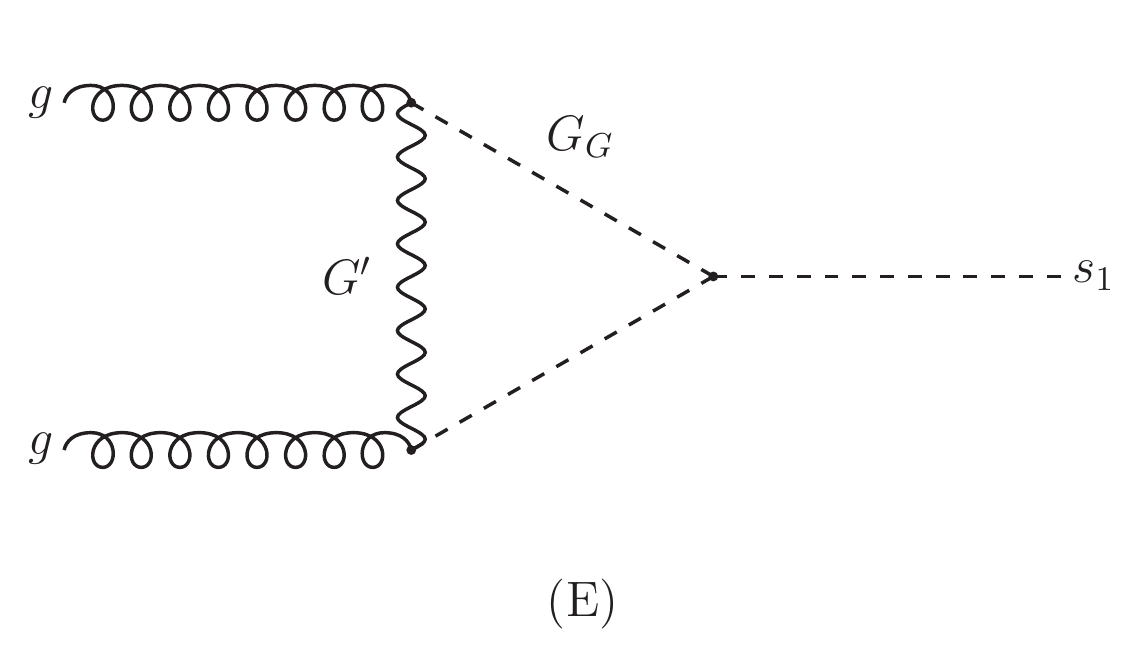}
\includegraphics[scale=0.45, angle=0]{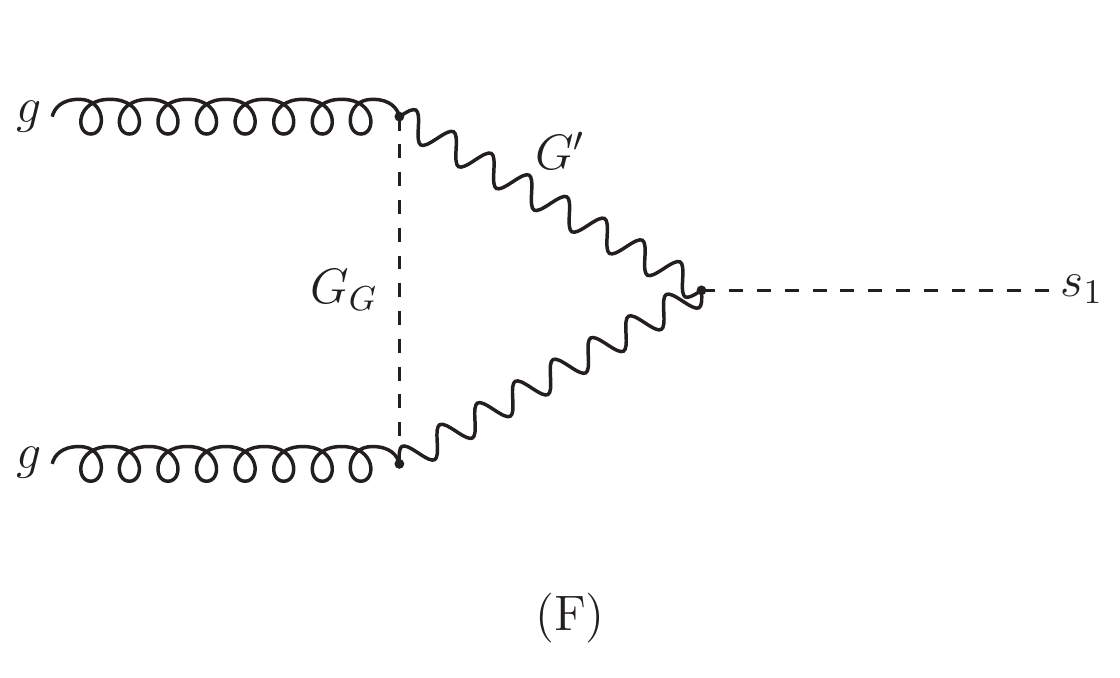}
\includegraphics[scale=0.45, angle=0]{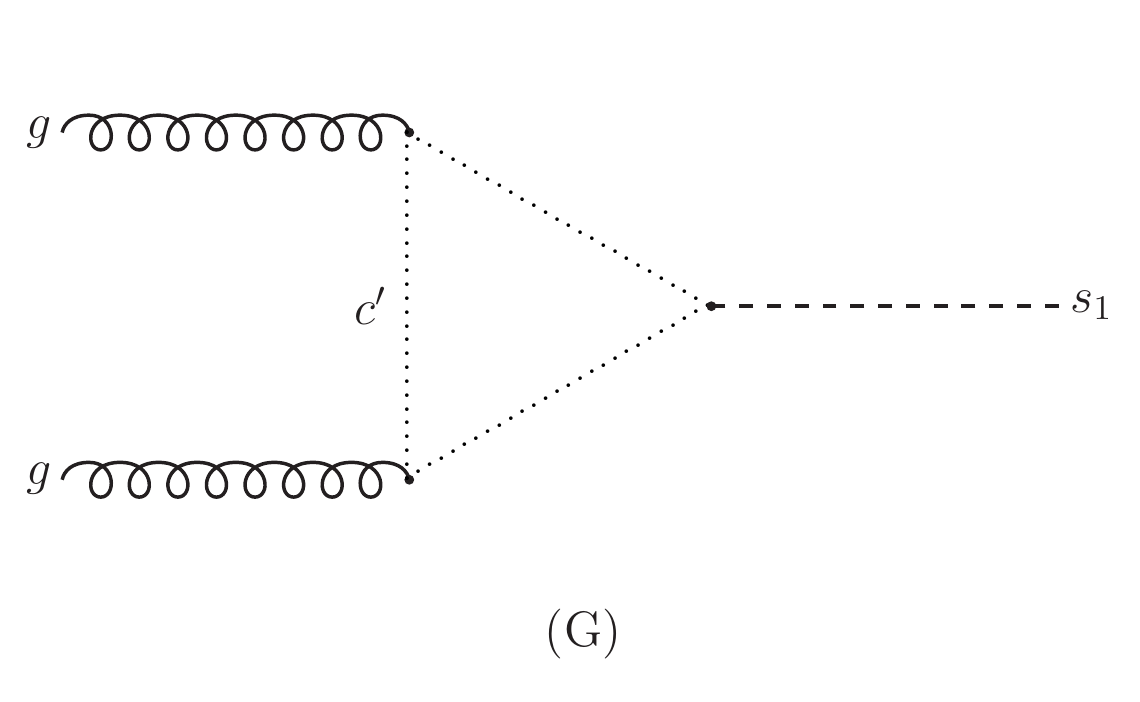}
\includegraphics[scale=0.45, angle=0]{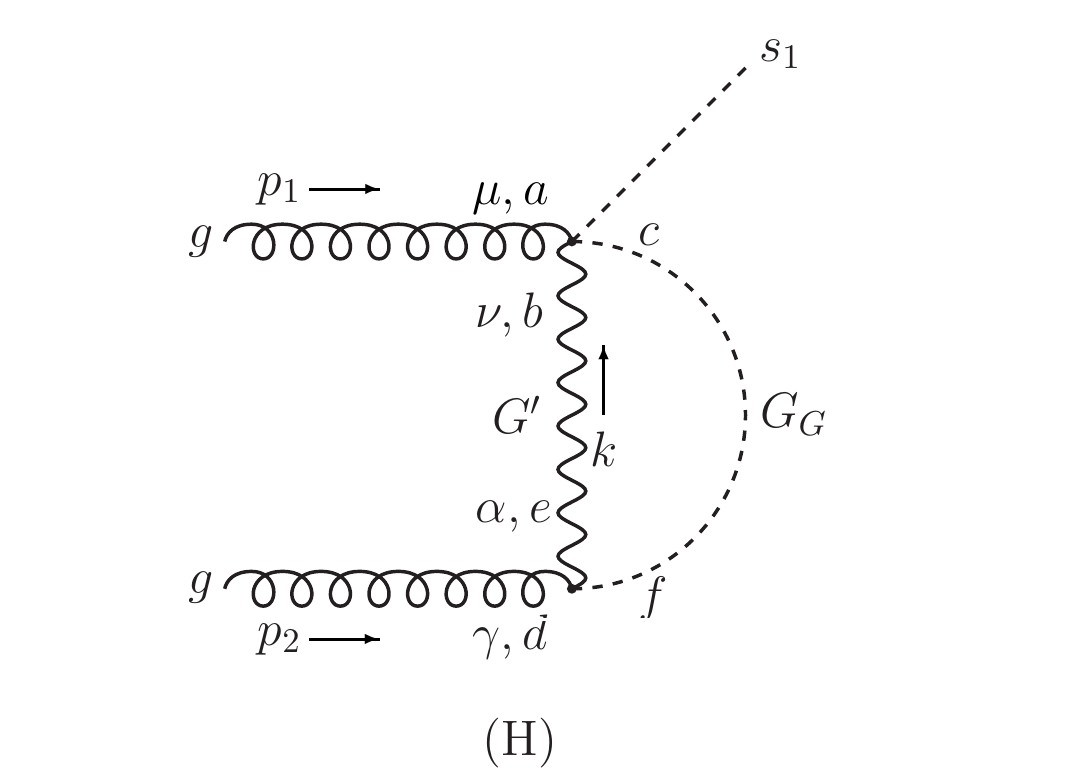}
\includegraphics[scale=0.45, angle=0]{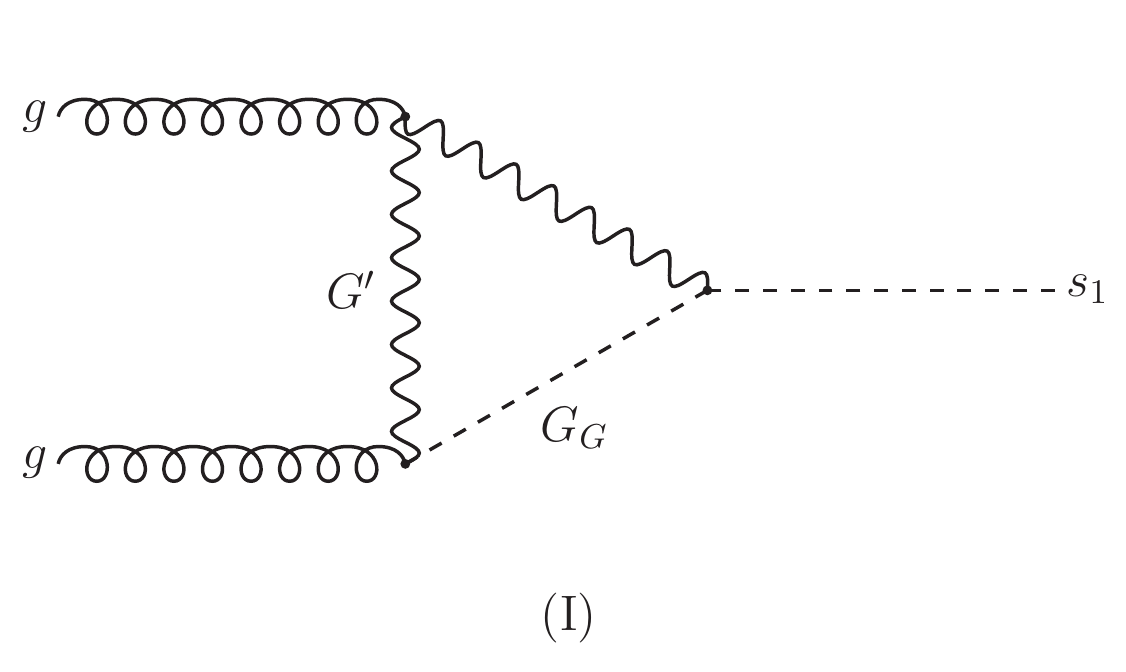}
\includegraphics[scale=0.45, angle=0]{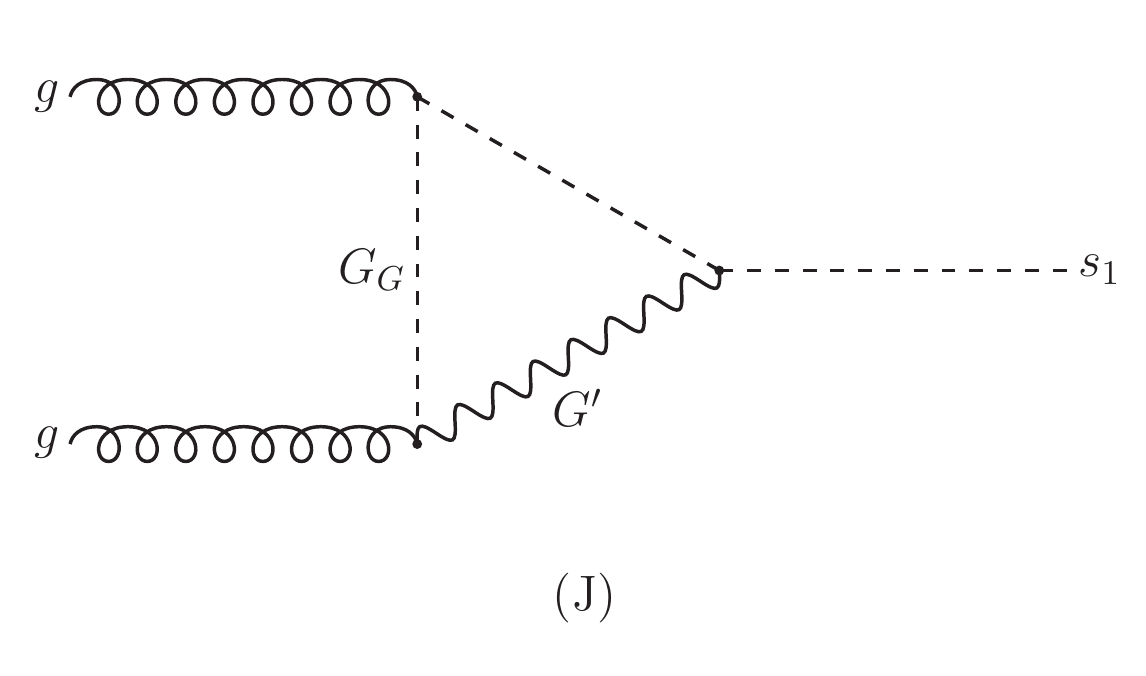}
\caption{\label{fig:Feynmanvectorloop}{Feynman diagrams for vector
    loop contributions to $gg \rightarrow s_1$ in the Feynman
    gauge.}}
\end{figure}
As a cross-check of our result in~\subsecref{Vector_unitary_explicit},
we perform the same calculation in Feynman gauge, setting $R_\xi = 1$.
Again, we adopt the ReCoM with the Higgs portal, as discussed
in~\subsecref{ReCoM}. In contrast to the calculation done in the
unitary gauge, here we perform the calculation in the scalar mass
basis, explicitly deriving the corresponding Goldstone couplings to
$s_1$ after taking into account the $h$--$\phi_R$ mixing.  This
motivates an interesting discussion of the coupling of Goldstone
bosons to their partner Higgs field when the partner Higgs field is
mixed with spectator scalars and provides a further check that Higgs
mixing and NP contributions to $s_{1,2}$ production can be factored as
in~\eqnref{ME_projections}.

\subsection{Goldstone couplings}
\label{subsec:GoldstoneSector}
Beginning with the full scalar potential
in~\eqnref{fullscalarpotential} and the exact vacuum expectation
values given in~\eqnsref{vPhi}{vH}, we examine the
Goldstone couplings to the scalars $h$ and $\phi$.  After expanding
the potential, we have
\begin{equation}
\label{eqn:Goldstone_Couplings}
V(\Phi) \supset \dfrac{1}{2}G_G^a G_G^d \delta^{ad}
\left( \phi_R \left( 
-\dfrac{\mu}{\sqrt{6}} + \dfrac{v_\phi}{3}(3\lambda_\Phi + \kappa_\Phi) \right)
+ h (-\lambda_H v_h) \right)
+ \dfrac{G_0^2}{2} 
\left( -\lambda_{hp} v_\phi \phi_R +  2\lambda_H v_h h \right) \ .
\end{equation}
This can be written in matrix form as,
\begin{equation}
\label{eqn:Goldstone_Couplings_MatrixForm}
\begin{array}{ccl}
V(\Phi) &\supset&
\left(
\begin{array}{cc}
\dfrac{h}{\sqrt{2}} & \dfrac{\phi_R}{\sqrt{2}}
\end{array} \right)
\left( 
\begin{array}{cc}
2\lambda_H v_h^2 & -\lambda_{hp} v_h v_\phi \\
-\lambda_{hp} v_h v_\phi & 
\dfrac{-\mu}{\sqrt{6}} v_\phi +
\dfrac{v^2_\phi}{3} (3\lambda_\Phi+\kappa_\Phi)
\end{array} \right)
\left(
\begin{array}{c}
\dfrac{G_0^2}{v_h\sqrt{2}} \\ 
\dfrac{G_G^a G_G^d \delta^{ad}}{v_\phi \sqrt{2}} \\ 
\end{array} \right) \\
&=& 
\left(
\begin{array}{cc}
\dfrac{h}{\sqrt{2}} & \dfrac{\phi_R}{\sqrt{2}}
\end{array} \right)
\hat{M}
\left(
\begin{array}{c}
\dfrac{G_0^2}{v_h\sqrt{2}} \\ 
\dfrac{G_G^a G_G^d \delta^{ad}}{v_\phi \sqrt{2}} \\ 
\end{array} \right) \ . \\
\end{array}
\end{equation}
Note if we set $\lambda_{hp} = 0$, then $v_h$ and $v_\phi$ become the
unperturbed vevs and the Goldstone couplings become $m_h^2 / v_h$ and
$m^2_{\phi_R}/v_\phi$ for the original $m^2_{\phi_R}$, $m_h^2$,
$v_\phi$, and $v_h$.  From~\eqnref{Goldstone_Couplings_MatrixForm}
and~\eqnref{mass_matrix_terms} we see that the
Goldstone--Goldstone--scalar interaction matrix $\hat{M}$ is the
same as the scalar mass matrix.  Thus when we diagonalize the mass
matrix, we will simultaneously diagonalize the Goldstone interaction
matrix in~\eqnref{Goldstone_Couplings_MatrixForm}.  Explicitly, we
write the scalar mass and Goldstone interaction terms as
\begin{equation}
\label{eqn:Mass_Plus_Goldstone_Lagrangian}
\mathcal{L} \supset
- \left( \begin{array}{cc}
\dfrac{h}{\sqrt{2}} & \dfrac{\phi_R}{\sqrt{2}}
\end{array} \right)
\hat{M}
\left( \begin{array}{c}
\dfrac{h}{\sqrt{2}} \\
\dfrac{\phi_R}{\sqrt{2}} 
\end{array} \right)
- \left( \begin{array}{cc}
\dfrac{h}{\sqrt{2}}  &  \dfrac{\phi_R}{\sqrt{2}}
\end{array} \right)
\hat{M}
\left( \begin{array}{c}
\dfrac{G_0^2}{v_h\sqrt{2}} \\ 
\dfrac{G_G^a G_G^d \delta^{ad}}{v_\phi \sqrt{2}} \\ 
\end{array} \right) \ .
\end{equation}
After diagonalization we obtain, for the interaction term,
\begin{equation}
\mathcal{L} \supset
- \left( \begin{array}{cc}
\dfrac{s_1}{\sqrt{2}} & \dfrac{s_2}{\sqrt{2}} \\
\end{array} \right)
\hat{M}_D \hat{U}^{-1}
\left( \begin{array}{c}
\dfrac{G_0^2}{v_h\sqrt{2}} \\ 
\dfrac{G_G^a G_G^d \delta^{ad}}{v_\phi \sqrt{2}} \\ 
\end{array} \right)
\end{equation}
where $\hat{M}_D = \hat{U}^{-1} \hat{M} \hat{U}$ is the diagonalized
mass matrix with eigenvalues given in~\eqnref{ms1_ms2}, and $\hat{U}$
is the unitary rotation matrix with its mixing angle defined
in~\eqnref{hphiR_mixing}.  We see that the Goldstone-Goldsone-$s_1$
interaction is
\begin{equation}
\label{GGphi}
\mathcal{L} \supset -s_\theta \dfrac{m_{s_1}^2}{v_\phi} s_1 G_G^a G_G^d
\delta^{ad} \ ,
\end{equation}
and so the Goldstone--Goldstone coupling to the scalar mass eigenstate
is proportional to the scalar mass squared and a mixing angle.  This
justifies our ability to factor out the Higgs mixing angle when
considering $s_{1,2}$ production.

\subsection{Continuation of the Feynman Gauge Calculation}
\label{subsec:Feynman_gauge_continued}

Returning to the calculation, there are ten diagrams in the Feynman
gauge that must be calculated, as shown
in~\figref{Feynmanvectorloop}.  Using $p_a = k + p_1$ and $p_b = k
-p_2$, $p_c = p_1 + p_2 - k$, ~\eqnref{three_vertex},
and~\eqnref{four_vertex}, we obtain the amplitudes
\begin{eqnarray}
\label{eqn:Feynman_ME}
\vspace{4pt} 
i{\mathcal M}^{ad}_A &=& 
-2\left( -s_\theta \dfrac{m_{G^\prime}^2}{v_\phi} \right) 
g_s^2 f^{abc} f^{dcb} \epsilon_{1\mu}
\epsilon_{2\gamma} \int \dfrac{d^dk}{(2\pi)^d} \dfrac{g_{\alpha \nu}
  g_{\beta \delta} g_{\sigma \rho} g^{\delta \sigma} V^{\mu\nu\rho}
  V^{\gamma\beta\alpha}}{
   (p_a^2 - m_{G^\prime}^2) (p_b^2 - m_{G^\prime}^2)(k^2 - m_{G^\prime}^2)} \ , \\
\vspace{4pt} 
i{\mathcal M}^{ad}_B &=& \left( \dfrac{1}{2} \right) 
\left( -s_\theta \dfrac{m_{G^\prime}^2}{v_\phi} \right)
\epsilon_{1\mu} \epsilon_{2\gamma} \int \dfrac{d^dk}{(2\pi)^d}
\left (-i g_{\rho \beta} \delta^{bc} \right) V_{acdb}^{\mu\rho\gamma\beta} 
\dfrac{1}{ (p_c^2 - m_{G^\prime}^2)(k^2 - m_{G^\prime}^2)} \ , \\
\vspace{4pt} 
i{\mathcal M}^{ad}_C &=& 
-g_s^2 f^{abc} f^{dcb} 
\left( -s_\theta \dfrac{m_{s_1}^2}{v_\phi}\right) 
\epsilon_{1\mu} \epsilon_{2\gamma} \int \dfrac{d^dk}{(2\pi)^d} 
\dfrac{ \left(p_a^\mu + k^\mu\right) \left(p_b^\gamma + k^\gamma\right)}{
\left(p_a^2-m_{G^\prime}^2\right) \left(k^2-m_{G^\prime}^2\right)
\left(p_b^2-m_{G^\prime}^2\right)} \ , \\
\vspace{4pt} 
i{\mathcal M}^{ad}_D &=& 
- g_s^2 f^{bde}f^{bae} 
\left( -s_\theta \dfrac{m_{s_1}^2}{v_\phi}\right) 
\epsilon_{1\mu} \epsilon_{2\gamma} g^{\mu\gamma}  \int \dfrac{d^dk}{(2\pi)^d} 
\dfrac{1}{(k^2-m_{G^\prime}^2)(p_c^2-m_{G^\prime}^2)} \ , \\
\vspace{4pt} 
i{\mathcal M}^{ad}_E &=& 
g_s^2 m_{G^\prime}^2 \left( -s_\theta \dfrac{m_{s_1}^2}{v_\phi}\right)f^{abc}f^{dcb}
g^{\mu \gamma} \epsilon_\mu \epsilon_\gamma \int
\dfrac{d^dk}{(2\pi)^d}
\dfrac{1}{(k^2-m_{G^\prime}^2)(p_a^2-m_{G^\prime}^2)(p_b^2-m_{G^\prime}^2)} \ , \\
\vspace{4pt} 
i{\mathcal M}^{ad}_F &=& 
2 g_s^2 m_{G^\prime}^2 \left( -s_\theta \dfrac{m_{G^\prime}^2}{v_\phi} \right) 
f^{abc}f^{dcb} g^{\mu \gamma}
\epsilon_\mu \epsilon_\gamma \int 
\dfrac{d^dk}{(2\pi)^4}
\dfrac{1}{(k^2-m_{G^\prime}^2)(p_a^2-m_{G^\prime}^2)(p_b^2-m_{G^\prime}^2)} \ , \\
\vspace{4pt} 
i{\mathcal M}^{ad}_G &=& 
2 g_s^2 \left( -s_\theta \dfrac{m_{G^\prime}^2}{v_\phi}\right) f^{abc} f^{dcb}
\epsilon_{1\mu} \epsilon_{2\gamma} \int \dfrac{d^dk}{(2\pi)^d}
\dfrac{k^\mu k^\gamma}{
 (p_a^2 - m_{G^\prime}^2)(p_b^2-m_{G^\prime}^2)(k^2 - m_{G^\prime}^2)} \ , \\
\vspace{4pt} 
i{\mathcal M}^{ad}_H &=& 
2 g_s^2 f^{abc} f^{dcb} g^{\mu \gamma} 
\left( -s_\theta \dfrac{m_{G^\prime}^2 }{v_\phi} \right)
\epsilon_{1\mu} \epsilon_{2\gamma} \int \dfrac{d^dk}{(2\pi)^d}
\dfrac{1}{ (k^2-m_{G^\prime}^2)(p_b^2-m_{G^\prime}^2)} \ , \\
\vspace{4pt} 
i{\mathcal M}^{ad}_I &=& 
-2 g_s^2 f^{abc} f^{dcb} \left( -s_\theta \dfrac{m_{G^\prime}^2}{v_\phi} \right) 
\epsilon_{1\mu} \epsilon_{2\gamma}
\int \dfrac{d^dk}{(2\pi)^d} \dfrac{g_{\alpha
    \nu} g_{\sigma \rho} g^{\alpha \gamma}
  V^{\mu\nu\rho} (k-p_1-2p_2)^\sigma}{ 
(p_a^2-m_{G^\prime}^2)(p_b^2-m_{G^\prime}^2)(k^2-m_{G^\prime}^2)} \ , \\
\vspace{4pt} 
i{\mathcal M}^{ad}_J &=& -2 g_s^2
f^{abc} f^{dcb} \left( -s_\theta \dfrac{m_{G^\prime}^2}{v_\phi} \right) 
\epsilon_{1\mu} \epsilon_{2\gamma}
\int \dfrac{d^dk}{(2\pi)^d}
\dfrac{g_{\beta \delta} g^{\beta \gamma} (k-p_2)^\delta
  (2k+p_1)^\mu}{ 
(p_a^2-m_{G^\prime}^2) (p_b^2-m_{G^\prime}^2)(k^2-m_{G^\prime}^2)} \ .
\end{eqnarray}
After converting to Feynman parameters, calculating the loop
integrals, summing all of the amplitudes, and some simplification, the
divergences cancel and we obtain for the production of $s_1$,
\begin{equation}
\label{eqn:Feynman_amplitude}
i{\mathcal M}^{ad}_V = -i s_\theta \left( \dfrac{\alpha_s}{\pi}  \right) 
\left(\dfrac{C(r_{G^\prime})}{4 v_\phi} \right) 
\delta^{ad} \epsilon_{1\mu} \epsilon_{2\gamma} 
\left( p_1^\gamma p_2^\mu - \dfrac{m_{s_1}^2}{2} g^{\mu\gamma} \right)
F_V(\tau_{G^\prime}) \ ,
\end{equation}
in agreement with~\eqnref{Unitary_amplitude} and~\eqnref{ME_projections}.

\end{appendix}

\bibliographystyle{apsrev}
\bibliography{ggH}

\end{document}